%% file: paper_IA.tex
\documentclass[useAMS,usenatbib]{mn2e}

\input{abbrev}

\input{psfig.sty}

\usepackage{graphicx}
\usepackage{epsfig}
\usepackage{amssymb}
\usepackage{amsmath}
\usepackage{amsfonts}
\usepackage{txfonts}
\usepackage{multirow}
\usepackage{afterpage}
\usepackage{xcolor}

\definecolor{mypink1}{rgb}{0.858, 0.188, 0.478}
\definecolor{mygreen1}{rgb}{0.258, 0.788, 0.878}
\definecolor{myorange1}{rgb}{0.5, 0.2, 0.2}

\newcommand{\snr}{$\mathcal{S}/\mathcal{N}$ }
\newcommand{\mycomment}[1]{}

\title[Lensing beyond 2pt: accounting for IA]{Cosmic shear beyond 2-point statistics: Accounting for galaxy intrinsic alignment with projected tidal fields}

\author[J. Harnois-D\'{e}raps et al.]{Joachim Harnois-D\'{e}raps$^{1}$\thanks{E-mail: jharno@roe.ac.uk}, Nicolas Martinet$^{2}$ \& Robert Reischke$^{3}$
\\
$^{1}$School of Mathematics, Statistics and Physics, Newcastle University, Herschel Building, NE1 7RU, Newcastle-upon-Tyne, UK\\
$^{2}$Aix-Marseille Univ, CNRS, CNES, LAM, Marseille, France\\
$^{3}$Ruhr University Bochum, Faculty of Physics and Astronomy, Astronomical Institute (AIRUB), \\ German Centre for Cosmological Lensing, 44780 Bochum, Germany\\
}

\date{Accepted XXX. Received YYY; in original form ZZZ}

\pubyear{2021}

\begin{document}
\label{firstpage}
\pagerange{\pageref{firstpage}--21}
\maketitle

\begin{abstract}
Developing analysis pipelines based on statistics beyond two-point functions is critical for extracting a maximal amount of cosmological information from current and upcoming weak lensing surveys. In this paper, we study the impact of the intrinsic alignment of galaxies (IA) on three promising probes measured from aperture mass maps -- the lensing peaks, minima and full PDF. Our two-dimensional IA infusion method converts the light-cone-projected mass sheets into projected tidal tensors, which are then linearly coupled to an intrinsic ellipticity component with a strength controlled by the coupling parameter $A_{\rm IA}$. We validate our method with the $\gamma$-2PCFs statistics, recovering well the linear alignment model of \citet{BridleKing} in a full tomographic setting, and for different $A_{\rm IA}$ values. We next use our method to infuse at the galaxy catalogue level a non-linear IA model that includes the density-weighting term introduced in \citet{Blazek2015}, and compute the impact on the three aperture mass map statistics. We find that large \snr peaks are maximally affected, with deviations reaching 30\% (10\%) for a {\it Euclid}-like (KiDS-like) survey. Modelling the signal in a $w$CDM cosmology universe with $N$-body simulations, we forecast the cosmological bias caused by unmodelled IA for 100 deg$^2$ of {\it Euclid}-like data, finding very large offsets in $w_0$ (5-10$\sigma_{\rm stat}$), $\Omega_{\rm m}$ (4-6$\sigma_{\rm stat}$), and $S_8 \equiv \sigma_8\sqrt{\Omega_{\rm m}/0.3}$ ($\sim$3$\sigma_{\rm stat}$). The method presented in this paper offers a compelling avenue to account for IA in beyond-two-point weak lensing statistics, with a flexibility comparable to that of current $\gamma$-2PCFs IA analytical models.
\end{abstract}




\section{Introduction}
\label{sec:intro}

Weak gravitational lensing by large scale structure, commonly referred to as `cosmic shear', has provided some of the most stringent constraints on the key cosmological parameters that describe the dark sector of our Universe  \citep{KiDS1000_Asgari, DESY3_Amon, HSCY1_2PCF, HSCY1_Cell}. Dedicated Stage-III survey such as  the Kilo Degree Survey\footnote{kids.strw.leidenuniv.nl}, the Dark Energy Survey\footnote{www.darkenergysurvey.org} and the HyperSuprime Camera Survey\footnote{hsc.mtk.nao.ac.jp/ssp/} are either complete or nearing completion, and so far mostly agree on the value of $S_8\equiv\sigma_8\sqrt{\Omega_{\rm m}/0.3}$, the structure growth parameter that is best measured from lensing. The latter is defined as   
a combination of the matter density $\Omega_{\rm m}$ and clumpiness $\sigma_8$, which specifies the amplitude of  density fluctuations in spheres of 8$h^{-1}$Mpc.

These cosmological constraints are primarily inferred from measurements of two-point statistics, either the lensing power spectrum or the shear two-point correlation functions ($\gamma$-2PCFs), which are powerful summary statistics that can be analysed with prediction models now reaching an accuracy of a few percent \citep{HMCode2020, EuclidEmulator, DarkEmulator}.
The high accuracy attained by the two-point function analyses come however with a large cost in precision. Indeed, these methods  completely overlook the  non-Gaussian information contained in the mode coupling and in the phase correlation \citep{ChiangColes2000}, which will become increasingly important for the upcoming generation (Stage-IV) of lensing surveys. This has led to the development of a number of alternative measurement techniques, among which the lensing peak count statistics has received a particularly large attention from the community, primarily for the simplicity of its methods and for its effectiveness at capturing additional information \citep{2015PhRvD..91f3507L, 2015MNRAS.450.2888L, Kacprzak2016, Martinet18, Shan18, HD20}. Other promising methods worth mentioning are the lensing PDF \citep[][sometimes referred to the one-point statistics]{Boyle2020, Martinet20}  and other moments of the convergence map \citep{VanWaerbeke2013, Gatti20}, lensing minima \citep{MinimaPeaks}, Minkowski functional \citep{2015PhRvD..91j3511P}, shear clipping \citep{Giblin18}, lensing by voids \citep{Davies20}, deep learning with Convolutional Neural Networks \citep{Fluri2019}, persistent homology analysis and scattering transform of the lensing field \citep[respectively][]{Homology, Scattering}; these all present an appealing potential at improving the parameter constraints.  For example the joint peaks/$\gamma$-2PCFs have been shown to increase by a factor two the constraints on  $S_8$, by a factor three the precision on the dark energy equation of state parameter $w_0$ \citep{Martinet20} and by  $\sim$40\% the precision on the sum of the neutrino masses \citep{MassiveNu1, MassiveNu2} for Stage-IV surveys, compared to the $\gamma$-2PCFs alone.

An important feature common to  many of these approaches is the absence of  analytical models with which to predict the observed signals. Consequently, the cosmology inference must be completely calibrated from mock surveys constructed from suites of numerical weak lensing simulations, such as those described in \citet{DH10}, or the more recent {\it MassiveNuS}\footnote{ columbialensing.org/\#massivenus} \citep{MassiveNuS}  and {\it cosmo}-SLICS\footnote{slics.roe.ac.uk} \citep{cosmoSLICS}.

Another key challenge faced by these techniques relates to the handling of systematic uncertainties that are known to exist in the lensing data \citep[see][for a review]{Mandelbaum18}, and that must therefore also be accounted for in these alternative measurement methods. In particular, the uncertainties associated to photometric redshifts, shape calibration/shear inference, baryonic feedback mechanism and intrinsic alignment (IA) of galaxies
must be carefully  dealt with for the inferred cosmology to be accurate. While the former two of these effects can be included in  $N$-body simulations at the ray-tracing level, the impact of baryon feedback on lensing statistics beyond two-point statistics typically requires to be calibrated on hydrodynamical simulations  \citep{Osato2015, LensingPDF_baryons, Martinet21}, although {\it baryonification} methods can also be of assistance \citep{Baryonification2,Weiss_Peaks_Baryons,LuHaiman21}. 

The impact of IA has been mostly studied in the context of weak lensing two-point statistics, with only a few attempts to propagate the effect onto alternative statistics\footnote{For example, \citet{Gatti20}  modelled the IA in lensing moment  via their dependence on the two-point functions.}. 
There are in fact a number of theoretical models that attempt to describe the physics of IA, including the Non-Linear tidal Alignment model \citep[NLA hereafter, see][]{BridleKing}, the tidal torquing model \citep[e.g.][]{Hirata2004,Catelan_IA_Tidal}, a combination of both \citep[the Tidal Alignment and Tidal Torquing model, TATT hereafter, described in][]{Blazek2019} or  halo-based alignment model \citep{SchneiderBridle2010, Fortuna2020}. These all produce predictions for two-point statistics, which generally depend on  galaxy type, redshift, cosmology, with a few free parameters  calibrated on  hydrodynamical simulations \citep[see, e.g.][]{Samuroff2020,Zjupa2020} and observations. For example, \citet{2013MNRAS.431..477J} detect and constrain an IA signal in the COSMOS early-type galaxies, but find hardly any signal  for late-type galaxies; the WiggleZ blue galaxies are shown in  \citet{BlueIA} to be consistent with no IA;  \citet{Singh_IA_LOWZ} find a significant IA signal in the BOSS LOWZ sample; a similar trend was found by \citet{Johnston_IA}, from the KiDS, SDSS and GAMA surveys, with no signal detection for the blue galaxies, but a $9\sigma$ detection for the red. When interpreted within in the NLA model, they measure an amplitude parameter of $A_{\rm IA} = 3.18_{-0.46}^{+0.47}$.  As reported in \citet{Samuroff2020}, the different hydrodynamical simulations do not all agree on the physical model that best describes the IA. For example, the tidal torquing model was shown by \citet{Zjupa2020} to be strongly disfavoured over the NLA model in the {\it IllustrisTNG} simulation\footnote{www.tng-project.org/about/}, while it was observed for high-redshift blue galaxies in the {\sc Horizon-AGN} simulations\footnote{horizon-simulation.org} \citep{Codis2015} and in many other hydrodynamical experiments \citep[see][and references therein]{Chisari_IA2}. 

There is thus a large uncertainty on the strength of the galaxy alignments, and as a primary contaminant to the cosmic shear signal, it can be modelled and marginalised over from lensing analyses, providing indirect measurements of effective IA parameters. For example, the DES-Y1 reports non-zero values for the NLA  model parameters $A_{\rm IA}$ (the amplitude) and $\eta_{\rm IA}$ (the redshift evolution)  of $1.3^{+0.5}_{-0.6}$  and $3.7^{+1.0}_{-2.3}$ \citep{DESY1_Troxel}, respectively. Similarly, the KiDS-1000 analysis of \citet{KiDS1000_Asgari} find values of $A_{\rm IA}$ in the range $0.264^{+0.424}_{-0.337} - 0.973^{+0.292}_{-0.383}$, depending on the choices of two-point statistics, while the HSC cosmic shear analyses report $A_{\rm IA}$ of  $0.38\pm0.70$ and $0.91^{+0.29}_{-0.38},$ also depending on the choice of statistics \citep{HSCY1_Cell, HSCY1_2PCF}. The recent analysis of DES-Y3 favours slightly negative values of $A_{\rm IA}$, both when analysed with the NLA and the TATT model \citep{DESY3_Secco}.

The effect of IA is significant, and can cancel over 10-100\% of the observed cosmic shear signal, depending on the angular scale and redshift samples. If left unmodelled, the DES-Y1 and DES-Y3 inferred cosmologies would be biased by $1\sigma$ \citep{DESY1_Troxel,DESY3_Secco}, an effect that would  worsen to 5-6$\sigma$ in Stage-IV surveys such as Vera C. Rubin observatory\footnote{lsst.org} or the {\it Euclid}\footnote{sci.esa.int/web/euclid} and Nancy Grace Roman\footnote{roman.gsfc.nasa.gov} space telescopes, as reported in \citet[][see their figure 14]{Joachimi_IA_review}. In fact,  \citet{Blazek2019} have shown that if the true physics of the IA is described by the TATT, but the cosmology inference was assuming the NLA model, an LSST-like survey would still be several $\sigma$ away from the truth in many $w$CDM parameters. 

However, it is still under debate how well the TATT model can describe the alignment process in general \citep{Zjupa2020}, since no alignment signal has been observed so far for blue galaxies. \citet{DESY1_IA_Samuroff} for example find an IA amplitude of $A_\mathrm{IA} = 2.38^{+0.32}_{-0.31}$ for early-types described by the NLA model, but an amplitude consistent with zero for late-type galaxies which should preferentially align with the torquing mechanism captured by the TATT model according to some hydrodynamical simulation results. With the next generation of surveys the situation will change since the majority of their galaxy sample are expected to be blue galaxies.

IA  are expected to be significant on alternative  lensing  statistics as well and must therefore be harnessed if we are to exploit  their increased statistical power  and interpret their results correctly. This topic has received very little attention so far, and it is this gap that we intend to fill in this paper.  Our approach relies on the infusion of a physically-motivated IA signal in mock lensing catalogues based on dark matter-only simulations. This has been explored to limited extend in the literature, either within a Halo Occupation Distribution model based on halo shapes  \citep[as in][]{Heymans2006, Joachimi2013}, or by reweighting mass shells in the ray-tracing simulations   \citep{Fluri2019, Zuercher2020a}, however the accuracy and flexibility of these  methods are not satisfactory given the current data, and certainly will not meet the requirements set for the upcoming Stage-IV lensing surveys. In particular, halo-based methods require an accurate measurement of the halo inertia matrix, which in turns requires hundreds of particles for its measurement and therefore places a low-mass cut on haloes for which an IA signal can be assigned. 

We opted here instead to infuse IA in a manner that is physically consistent with the NLA model, e.g. based on a linear coupling between the intrinsic galaxy shapes and the non-linear projected tidal fields. We further include the first TATT term,  a density-weighting component that increases the impact of IA on small scales with respect to the NLA. Our method has a number of advantages over the previous studies. First, an intrinsic shape is given to every galaxy, allowing for easy catalogue-based analyses for any choice of lensing statistics, with the possibility to dissect the underlying true IA contribution.  We can also correctly combine the ellipticities and compute the reduced shear, which is the true lensing observable. Second, the coupling strengths are two free parameters that can be directly associated with the NLA $A_{\rm IA}$ and the TATT $b_{\rm TA}$ parameters, such that we can validate our infusion methods with two-point theoretical models. This connection further allows for joint analyses between two-point functions and higher-order statistics, in which the IA parameters could be  coherently varied and marginalised over. Third, our method only requires the projected mass sheets, which are generally stored by default for most weak lensing simulations, meaning that it can be computed straight-forwardly at  different  cosmologies from the public simulations suites mentioned above  in order to investigate possible degeneracies. Fourth, being based on the projected tidal field, it has the flexibility to adapt to many IA models, including those involving higher moments of the tidal field and/or local coupling between the tidal and density fields such as TATT. 

This paper is structured as follow. We first briefly review in Sec. \ref{sec:theory} the theory of weak lensing and intrinsic alignments, and describe in Sec. \ref{sec:sims} our weak lensing simulations and IA infusion models. We validate in  Sec. \ref{sec:validation} our IA-infused mocks against the theoretical NLA predictions at the level of correlation functions, in the context of Stage-III lensing surveys. As a first demonstration of our methodology, we 
measure from the same galaxy catalogues the impact on peak statistics  in Sec. \ref{sec:Peaks}, then explore the dependencies on smoothing scales and coupling strength. We next investigate the  cosmological biases caused by IA for different analysis designs in the context of Stage-IV lensing surveys.
More precisely, we carry out likelihood analyses based on three aperture mass maps statistics -- lensing peaks, lensing minima and lensing PDF --  and their combination with the shear correlation functions, measuring in each case the impact on the inferred $w$CDM parameters $S_8$, $\Omega_{\rm m}$ and $w_0$. We finally discuss our results and present our conclusions  afterwards, in Sec. \ref{sec:conclusion}.

\section{Theory and Modelling}
\label{sec:theory}

As mentioned in the introduction, there exist robust predictions for two-point statistics, however analytical modelling is highly limited when it comes to alternative measurement methods. This section briefly reviews the theory behind the modelling of cosmic shear correlation function and of the IA signal.

\subsection{Cosmic shear 2-point functions}
\label{subsec:wl-th}

The cosmological information contained in cosmic shear data on large linear scales is well captured by two-point statistics, either in the form of the lensing power spectrum $C_{\ell}$ or the shear two-point correlation functions  $\xi_{\pm}(\vartheta)$ ($\gamma$-2PCFs hereafter).
Both are related to the matter power spectrum  $P_{\delta}(k,z)$, for which accurate prediction tools such as {\sc Halofit} \citep{Smith03, Takahashi2012}, {\sc HMcode} \citep{HMCode2020}, the {\sc DarkEmulator} \citep{DarkEmulator} or the {\sc EuclidEmulator} \citep{EuclidEmulator} exist. This connection largely explains why the two-point statistics stand as a natural choice for compressing and interpreting the weak lensing data. In the Limber approximation, the lensing power spectrum between  tomographic bins `$i$' and `$j$'  is computed as:
 \begin{eqnarray}
C_{\ell}^{ij} = \int_0^{\chi_{\rm H}}  \frac{q^i(\chi) \,q^j(\chi)}{\chi^2} \, P_{\delta}\, \bigg(\frac{\ell+1/2}{\chi},z(\chi)\bigg) \ {\rm d}\chi,
\label{eq:C_ell}
\end{eqnarray}
where $\chi_{\rm H}$ is the co-moving distance to the horizon. The  lensing kernels $q^{i}$ depend on the redshift distribution $n(z)$ as:
\begin{eqnarray}
q^i(\chi) = \frac{3}{2}\Omega_{\rm m} \, \bigg(\frac{H_0}{c} \bigg)^2 \frac{\chi}{a(\chi)} \int_{\chi}^{\chi_{\rm H}} n^i(\chi')\frac{{\rm d}z}{{\rm d} \chi'}\frac{\chi' - \chi}{\chi'}{\rm d}\chi',
\label{eq:q_lensing}
\end{eqnarray}
where $c$ is the speed of light, $a$ is the scale factor and $H_0$ the Hubble parameter, note that $\chi_\mathrm{H} = c/H_0$.
Predictions for the $\gamma$-2PCFs are computed from Eq. (\ref{eq:C_ell}) as:
 \begin{eqnarray}
\xi_{\pm}^{ij}(\vartheta) = \frac{1}{2\pi}\int_0^{\infty} C_{\ell}^{ij} \, J_{0/4}(\ell \vartheta) \, \ell \, {\rm d}\ell,
\label{eq:xipm_th}
\end{eqnarray}
where $J_{0/4}(x)$ are Bessel functions of the first kind.
We adopt in this paper the \citet{Takahashi2012}  {\sc Halofit} model when computing $P_{\delta}(k,z)$ and use the cosmological parameter estimation code {\sc cosmoSIS}\footnote{bitbucket.org/joezuntz/cosmosis/wiki/Home} \citep{cosmosis} to produce our  $\xi_{\pm}^{ij}$ predictions.

 \subsection{Non-linear alignment model}
 \label{subsec:IA_th}
 
The observed ellipticity of a galaxy ${\boldsymbol \epsilon}_{\rm obs} $ is a combination of its intrinsic shape ${\boldsymbol \epsilon}_{\rm int}$ and a cosmic shear signal ${\boldsymbol \gamma}$, the former of which can be further divided in a random component ${\boldsymbol \epsilon}^{\rm ran}$ and an alignment term ${\boldsymbol \epsilon}^{\rm IA}$.
According to  the NLA model\footnote{The term `non-linear' in the model name refers to the use of the non-linear matter power spectrum $P(k)$ in the calculations; the coupling with the tidal field is still linear.},  IA are caused by a linear coupling between galaxy shapes and the non-linear large-scale tidal field at the galaxy position.

In the context of  two-point function analyses, these intrinsic shapes contribute to an intrinsic-intrinsic ($II$) term as well as to an intrinsic-shear coupling ($GI$) term \citep{Hirata2004}, both secondary signals to the true cosmic shear ($GG$) term.  The $II$ and $GI$ terms  can be computed from the matter power spectrum as: 
 \begin{eqnarray}
P_{II}(k,z) =  \left(\frac{A_{\rm IA}\bar{C_1}\bar{\rho}(z)}{\overline{D}(z)}\right)^2a^4(z) P_{\delta}(k,z)
\label{eq:Pk_II_th}
\end{eqnarray}
and
 \begin{eqnarray}
P_{GI}(k,z) = - \frac{A_{\rm IA}\bar{C_1}\bar{\rho}(z)}{\overline{D}(z)}a^2(z) P_{\delta}(k,z) \, ,
\label{eq:Pk_GI_th}
\end{eqnarray}
which can then be Limber-integrated (as in Eqs. \ref{eq:C_ell} and \ref{eq:xipm_th}) to compute  the secondary signals $C_{\ell}^{II}$,  $C_{\ell}^{GI}$ and $\xi_{\pm}^{II}(\vartheta)$, $\xi_{\pm}^{GI}(\vartheta)$. 
Here  $\overline{D}$ is the `rescaled linear growth factor' defined as $\overline{D} \equiv D(1+z)$, $\bar{\rho}(z)$ is the matter density at redshift $z$ and $\bar{C_1}$ is a constant calibrated in \citet{Brown2002} that takes on the default value of $5\times 10^{-14} M_{\odot}^{-1} h^{-2} {\rm Mpc}^3$. The amplitude parameter $A_{\rm IA}$ describes the strength of the tidal coupling and is the main NLA parameter constrained by current cosmic shear surveys. When phrased in terms of intrinsic ellipticities and tidal field $s_{ij}$,  the scale factors from the previous equations exactly cancel the redshift evolution of the matter density field, leaving:
 \begin{eqnarray}
\epsilon_1^{\rm IA} = - \frac{A_{\rm IA}\bar{C_1}\bar{\rho}(z=0)}{D(z)} (s_{xx} - s_{yy}) \,\,\,\,,   \epsilon_2^{\rm IA} = - \frac{2 A_{\rm IA}\bar{C_1}\bar{\rho}(z=0)}{D(z)} s_{xy}\;,
\label{eq:tidal_th}
\end{eqnarray}
where $s_{ij} = \partial_{ij}\phi$ are the Cartesian components of the tidal tensor of the gravitational potential $\phi$. Note that it is now the standard growth factor that appears in the denominator.

As discussed later, the key quantity of interest to cosmic shear analyses is not the tidal tensor itself but its trace-free version, since shape correlations with the density field are subdominant, even though peaks in high density environments are less elliptical on average. Consequently, the model itself has a restricted range of validity: on small scales, higher order couplings to the ellipticity field ${\boldsymbol \epsilon}^{\rm IA}$  become important, however these are neglected in the NLA model. Only the non-linear evolution of the tidal tensor itself is taken into account, since it does fit observations quite well.

We note that the NLA predicts additional higher-order terms and non-zero $B$-modes \citep{Hirata2004} that we neglect  in this analysis. Also, as shown in \citet{IA_EFT}, the NLA model can be interpreted as the lowest order description of the alignment process of galaxies in the light of an effective field theory description.

\subsection{Extension: $\delta$-NLA}
\label{subsec:IA_th_ext}

As mentioned in \citet{Blazek2015} and  \citet{Blazek2019}, the intrinsic alignment of galaxies can only be observed at the galaxy positions, which are not randomly distributed on the sky but instead trace the underlying matter density `$\delta$' with some biasing scheme. Accounting for this can be done at the level of theoretical predictions with a density weighting term computed in the above-mentioned references from perturbation theory, and at the mock-infusion stage, by imposing (or) not our mock galaxies to trace the underlying matter field. What matters for a good match is that the simulated galaxy catalogues are being analysed with a consistent model. In presence of this sampling effect, the observed ellipticities from Eq. (\ref{eq:tidal_th}) become
 \begin{eqnarray}
\epsilon_{1/2}^{\rm IA, \delta} = \epsilon_{1/2}^{\rm IA}\times(1 + b_{\rm TA}\delta).
\label{eq:tidal_th_deltaNLA}
\end{eqnarray}
The term $b_{\rm TA}$ encodes the coupling strength with the local density field, and is often set to unity but could be allowed to vary. Note that the standard NLA described in last section does not include this term, hence we refer to this improvement as the $\delta$-NLA model. The enhancement is mostly seen on small scales \citep{Blazek2019}, and has also been observed in hydro simulations \citep{Hilbert_IA2017}.
This is fully consistent with the TATT model in which the  torque term \citep[$C_2$ or $A_2$, see][]{Blazek2019,DESY3_Secco} is nulled, and further reduces to the NLA described in the last section by setting $b_{\rm TA}$ to zero.

\section{Simulations}
\label{sec:sims}

In this section we briefly review  the back-bone simulated light-cones, then  describe how we construct weak lensing catalogues infused with IA terms consistent with the NLA and $\delta$-NLA models. We also detail the four weak lensing estimators on which the impact of IA are measured (with results presented in Secs. \ref{sec:validation} and \ref{sec:Peaks}).

\subsection{Weak lensing light-cones} 
\label{subsec:WL_sims}

\begin{figure}
\begin{center}
\includegraphics[width=3.0in]{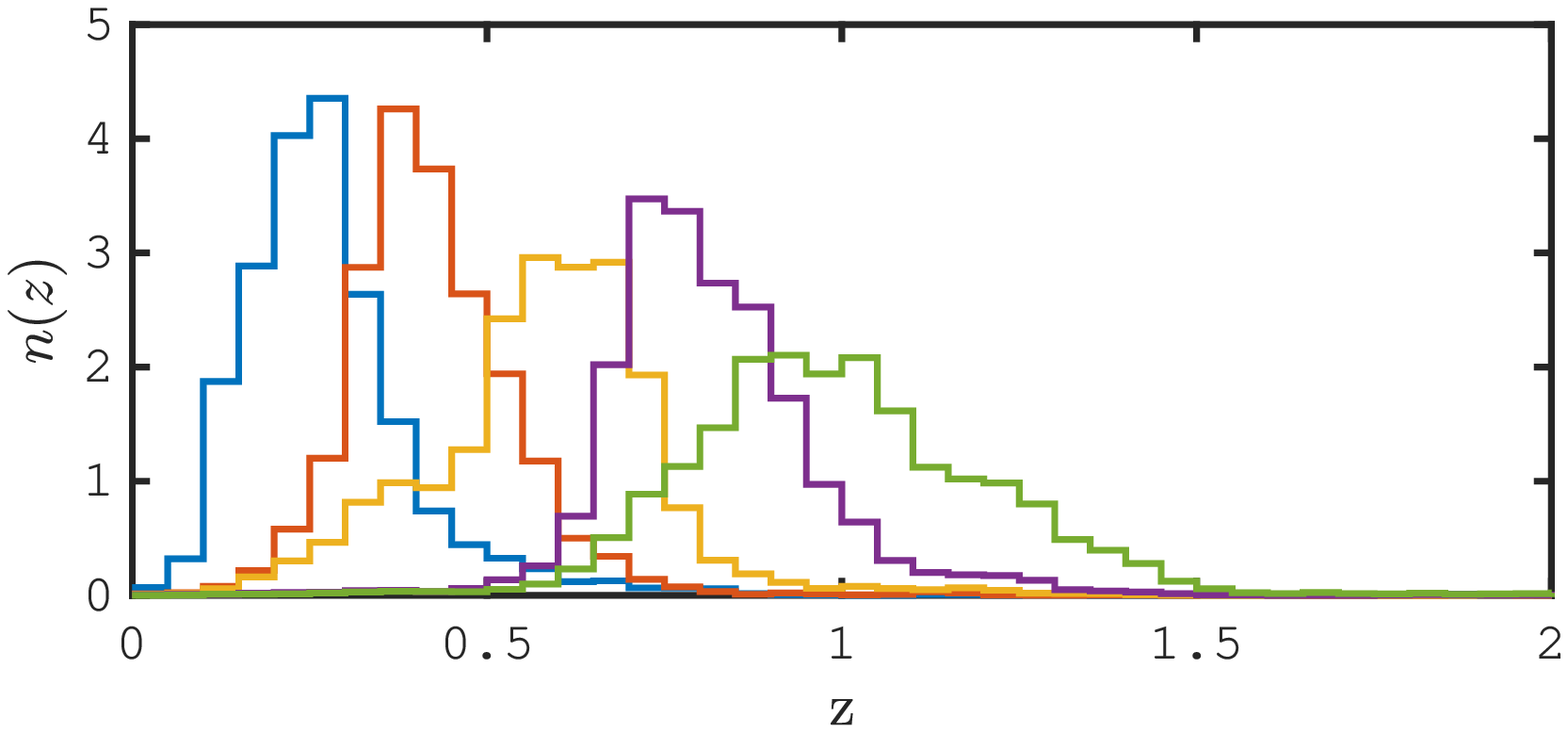}
\includegraphics[width=3.1in]{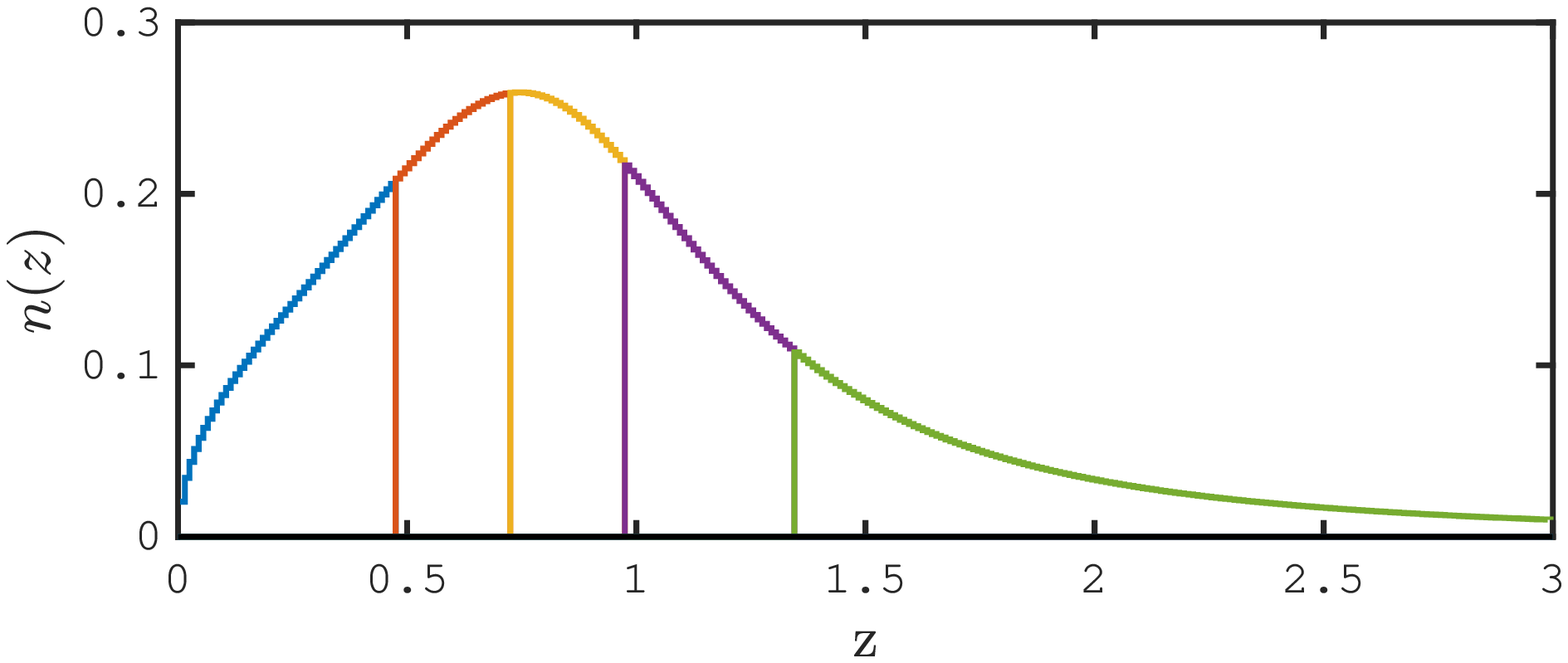}
\caption{Redshift distribution of the five tomographic bins, measured from the KiDS-like ({\it upper}) and {\it Euclid}-like  ({\it lower}) simulations. Note the change in the axes scaling between the two panels arising from different bin widths. In both panels, the  $n(z)$ distributions for the different tomographic bins are normalised such that $\sum n(z) {\rm d}z=1.0$.}
\label{fig:nz}
\end{center}
\end{figure}

Our simulated cosmic shear data are based on the Scinet LIght-Cone Simulations \citep[SLICS hereafter, see][]{SLICS} and the {\it cosmo}-SLICS \citep{cosmoSLICS}, which are two complementary  $N$-body suites designed for the analysis of Stage-III weak lensing data.  The SLICS are a series of 928 fully independent runs at a fixed flat $\Lambda$CDM cosmology ($\Omega_{\rm m}$=0.2905, $\sigma_8$=0.826, $\Omega_{\rm b}$ = 0.0474, $h$=0.6898 and $n_{\rm s}$=0.969), each evolving 1536$^3$ matter particles in a box of side $L_{\rm box}$=505$h^{-1}$Mpc. These simulations are specifically tailored for estimating weak lensing covariance matrices, and were central to a number of cosmic shear \citep[e.g.][]{KiDS450,Martinet18,Giblin18,HD20} and combined-probe \citep{2016MNRAS.460..434H, 2017MNRAS.471.1619H,Joudaki18, vanUitert18, Brouwer18} data analysis. The {\it cosmo}-SLICS, in contrast, cover a wide range of $w$CDM parameter values, more specifically $\Omega_{\rm m}$, $S_8$, $h$ and $w_0$. They sample 25 points organised in a Latin hypercube in order to minimise the interpolation error, each evolving a pair of $N$-body runs designed to suppress the sampling variance. One of these nodes, the fiducial model, lies at a $\Lambda$CDM cosmology identical to that of the SLICS, albeit with a $\sigma_8$ value that is 1.2\% higher.

Every $N$-body simulations produced  sequences of projected density fields chosen such as to fill the past light-cones up to $z=3$ in steps of $L_{\rm box}/2$, with randomised origins and projection axes.  These $10\times10$ deg$^2$ mass maps, which we  label $\delta_{2D}(\boldsymbol \theta)$, are used  to produce multiple  convergence and shear maps, $\kappa(\boldsymbol \theta)$ and $\gamma_{1/2}(\boldsymbol \theta)$, from which lensing quantities can be interpolated at a given galaxy redshift and position. We refer the interested reader to \citet{SLICS} for complete details on how this is achieved.

\subsection{Weak lensing catalogues} 
\label{subsec:WL_cats}

\begin{table}
   \centering
   \caption{Properties of our KiDS-like and {\it Euclid}-like surveys.  For the former survey, the effective number densities $n_{\rm eff}$ [in gal/arcmin$^{2}$], shape noise per component $\sigma_\epsilon$ and redshift distributions listed here match those of the KiDS-1000 data presented in  \citet{KiDS1000_Giblin} and \citet{KiDS1000_redshifts}. The column `$Z_B$ range' refers to the photometric selection that defines the five KiDS-1000 tomographic bins, while the mean redshift in each bin is listed under $\langle z \rangle$. The specifications of the {\it Euclid}-like survey follows those presented in \citet{Martinet20}, with $n_{\rm eff}$=6.0 gal/arcmin$^2$ for each tomographic bin and $\sigma_\epsilon$=0.26 per component.}
   \tabcolsep=0.11cm
      \begin{tabular}{@{} lcccc|cc @{}} 
                  & \multicolumn{4}{c}{KiDS-like} &  \multicolumn{2}{c}{{\it Euclid}-like}\\
      \hline
      \hline
      tomo    & $Z_B$ range     &    $n_{\rm eff}$ & $ \sigma_\epsilon$  & $\langle z \rangle$ &$z$ range&  $\langle z \rangle$  \\
       \hline
       bin1    & $0.1 - 0.3$ &         0.616&0.270 & $0.257$& $0.0 - 0.4676$&$0.286$ \\
       bin2    & $0.3 - 0.5$ &         1.182& 0.258 & $0.402$& $0.4676 - 0.7194$& $0.600$ \\
       bin3    & $0.5 - 0.7$ &          1.854& 0.273 & $0.563$& $0.7194 - 0.9625$& $0.841$\\
       bin4    & $0.7 - 0.9$ &        1.259& 0.254 & $0.792$& $0.9625 - 1.3319$ &$1.134$\\
       bin5    & $0.9 - 1.2$ &        1.311& 0.270 & $0.984$& $1.3319 - 3.0$& $1.852$\\
    \hline 
    \hline
    \end{tabular}
    \label{table:survey}
\end{table}

As mentioned in the introduction, we investigate in this paper the impact of IA for two different surveys configurations.  We first construct mock datasets that resemble the Fourth Data Release of the Kilo Degree Survey, with shapes noise, galaxy density and redshifts statistically matching the data  properties described in \citet{KiDS1000_Giblin} and \citet{KiDS1000_redshifts}, respectively. These KiDS-like simulations are split in five tomographic bins, with redshift distributions shown in the upper panel of Fig. \ref{fig:nz} and key survey properties listed in Table \ref{table:survey}. Contrarily to the public SLICS and {\it cosmo}-SLICS mock data, in this work the positions of the mock galaxies are not placed at random, but assigned such as to trace the projected matter density with a linear galaxy bias set to unity. More precisely, the galaxies in each catalogue are ranked according to their redshift, they are next associated with the mass sheet $\delta_{2D}(\boldsymbol \theta)$  that is the closest in redshift, then assigned a position by sampling $\delta_{2D}(\boldsymbol \theta)$  \citep[as described in Appendix A2 of][]{SLICS}. 
This extra level of realism better reproduces the clustering properties seen in the data, but, more importantly for the current study, it also up-weights the number of galaxies that live in dense environment and that therefore experience strong tidal fields. Overlooking this aspect would result in an underestimation of the importance of IA, and is the key difference between the NLA and the $\delta$-NLA model \citep{Blazek2019}. These  mock data have a relatively low signal-to-noise per tomographic bin, hence we produce 50  IA-infused  KiDS-like catalogues from as many {\it cosmo}-SLICS light-cones, at the fiducial cosmology. They primarily serve to establish the validity of our infusion method with $\gamma$-2PCFs in this paper, but are also ideally suited for including IA modelling in ongoing cosmic shear analyses beyond two-points statistics. As an example, we examine the effect of IA on the peak count statistics measured from our mock catalogues, which will serve in an upcoming  analysis of the actual KiDS DR4 data.

We next assemble a series of {\it Euclid}-like mock galaxy catalogues specifically designed for assessing the impact of IA on Stage-IV cosmic shear measurements, from the summary statistics all the way to the cosmological inference stage.
For this purpose, we exploit the 928 SLICS and the 26$\times$10 {\it cosmo}-SLICS light-cones introduced in \citet{Martinet20} to estimate the covariance matrix and to model the $w$CDM cosmological dependence of our estimators, respectively\footnote{The galaxy positions were  assigned at random in these $w$CDM and covariance simulations, and as opposed to the density-weighted positions used in the IA-infused mocks. Since we measure the relative impact of IA relative to the density-weighted catalogues with $A_{\rm IA}$ set to 0.0, this systematic difference should not impact the inferred cosmology.}. We further construct a series of IA-infused mocks following the same method as for the KiDS-like catalogues, except that the redshift distribution is now given by:
\begin{eqnarray}
n(z) =A \frac{z^a + z^{ab}}{z^b + c} \,\, , 
\label{eq:nz_euclid}
\end{eqnarray}
with $A$=1.7865, $a=0.4710$, $b$=5.1843, $c$=0.7259, and a galaxy density of $n_{\rm gal}=30$ gal arcmin$^{-2}$. The catalogues are next  divided evenly across five tomographic bins, for a density of  $n_{\rm gal}=6.0$ gal arcmin$^{-2}$ per bin. The shape noise is set to $\sigma_{\epsilon} = 0.26$ per component (see Table \ref{table:survey} for a summary of these properties).  The resulting  $n(z)$ is presented in the lower panel of Fig. \ref{fig:nz}.
With a lower  level of shape noise  and a significantly higher galaxy density compared to the KiDS-like catalogues, \citet{Martinet21} find that 10 light-cones per cosmology is enough for their measurement (an analysis similar to ours but in the context of baryonic feedback).  For all {\it Euclid}-like catalogues, five shape noise realisations are produced and averaged over in the end, to better capture the non-Gaussian signal.

\subsection{Infusion of intrinsic alignments}
\label{subsec:IA_sims}

\begin{figure}
\begin{center}
\includegraphics[width=3.3in]{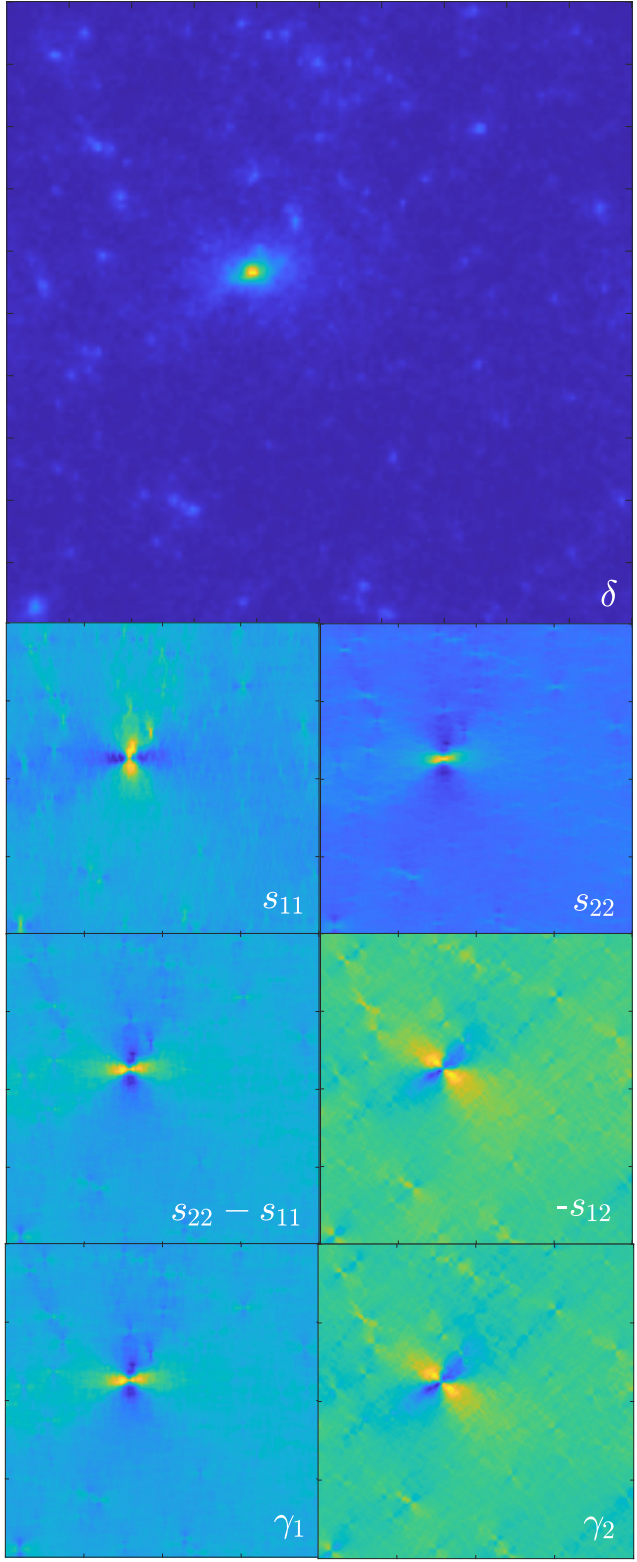}
\caption{({\it top:}) Projected mass over-density, measured at $z = 0.04$ at our fiducial cosmology,  subtending 2.6 deg on the side. ({\it second and third row:}) Different components of the tidal field tensor computed from the over-density map shown on the top (see Eq. \ref{eq:sij_2D}). ({\it lower row:}) The two lensing shear components produced by the same mass distribution, assuming a galaxy source plane at  $z_{\rm s} = 0.08$, i.e.  just behind.}
\label{fig:tidalator}
\end{center}
\end{figure}

The next step consists in computing the projected tidal fields within our simulations, which can then be coupled to an intrinsic  galaxy shape according to our IA models.
Given a three-dimensional mass over-density field $\delta(\boldsymbol x)$ and a smoothing scale $\sigma_{\rm G}$,  the trace-free tidal tensor $s_{ij}(\boldsymbol x)$ can be computed as \citep{Catelan_IA_Tidal}:
 \begin{eqnarray}
 \widetilde{s}_{ij} (\boldsymbol k)  = \left[\frac{k_i k_j}{k^2} - \frac{1}{3}\right]  \widetilde {\delta}(\boldsymbol k) \mathcal{G}(\sigma_{\rm G})
 \label{eq:sij}
\end{eqnarray}
where tilde symbols ` $\,\widetilde{ }\,$ '  denote Fourier transformed quantities,  the indices $(i,j)$ label the components of the Cartesian wave-vector $\boldsymbol{k}^T = (k_x,k_y,k_z)$, and $k^2 = k_x^2 + k_y^2 + k_z^2$. 
The smoothing kernel $\mathcal{G}(\sigma_{\rm G})$ is typically a three-dimensional Gaussian function described by a single parameter $\sigma_{\rm G}$, which determines what physical scales are allowed to affect the IA term in our model. We show in Appendix \ref{sec:3D_2D} (see Eq. \ref{eq:sij_2D_app}) that the components of the projected tidal shear $s_{ij,2D}(\boldsymbol \theta)\equiv \sum_{z} s_{ij} ({\boldsymbol x})$  can be computed from the projected density $\delta_{2D}(\boldsymbol \theta)$ as:
 \begin{eqnarray}
\widetilde{s}_{ij, 2D} (\boldsymbol k_{\perp})   = \sum_{z}  \widetilde{s}_{ij} ({\boldsymbol k}) =2\pi \left[\frac{k_i k_j}{k^2} - \frac{1}{3}\right]\widetilde{\delta}_{2D}(\boldsymbol k_{\perp}) \mathcal{G}_{2D}(\sigma_{\rm G}) \, ,
 \label{eq:sij_2D}
 \end{eqnarray}
where $\boldsymbol k_{\perp}$ represents the two Fourier components perpendicular to the line of sight, the indices ($i,j$) hereafter refer to either the $x$ or the $y$ component, and $\mathcal{G}_{2D}$ is now a two-dimensional smoothing kernel. An important aspect of our model is that neither the functional form of  $\mathcal{G}_{2D}(\sigma_{\rm G})$ nor the value of $\sigma_{\rm G}$ are specified within the NLA model. Smoothing amounts to selecting physical scales that do not contribute to the alignment of galaxies, which is still a debatable quantity, but is also introduced for numerical stability. \citet{Blazek2015} argues that 1.0 $h^{-1}$Mpc could be a reasonable fiducial value, being larger then the typical halo size, but recognizes that one-halo terms are also required to better match the observations. In their later work, \citet{Blazek2019} do not include smoothing at all, and neither do the KiDS-1000 nor DES-Y1 analyses based on the NLA model \citep{KiDS1000_Asgari,DESY1_Troxel}. In this work we used a two-dimensional Gaussian filter and calibrated $\sigma_{\rm G}$ empirically to $\sigma_{\rm G}=0.1h^{-1}$Mpc (see Sec. \ref{subsubsec:sigma_G}), however these two choices are arbitrary and could possibly be better optimised in the future; we also explore $\sigma_{\rm G}=0.5h^{-1}$Mpc later on.  We note that the smoothing scale is degenerate with the resolution of the simulation itself, and that one should use caution when smoothing on scales that approach the resolution limits.

We employ a numerical technique worth mentioning here: the Fourier transforms involved in computing Eq. (\ref{eq:sij_2D}) are computed from the full (projected) periodic boxes of the simulations, then interpolated on the light-cones. We find that tidal field computed directly from the $10\times10$ deg$^2$ light-cones  suffer from important large scales features, largely caused by  the non-periodic boundary conditions, which our method avoids.

The projected tidal field maps $s_{11}$, $s_{22}$ and $s_{12}$ are constructed   for each of the 18 mass sheets in every light-cones and interpolated at the position of every galaxy. Fig. \ref{fig:tidalator} illustrates this process for a small zoomed-in patch at $z\sim0$, starting from a $\delta_{2D}(\boldsymbol \theta)$ map (upper large panel), computing the tidal field components (middle four panels) and comparing the results with the cosmic shear signal generated by the same mass distribution (bottom two panels,  see Eq. \ref{eq:gamma_12}). The tidal field maps clearly reproduce the cosmic shear maps, and the minus sign in front of both terms in Eq. (\ref{eq:tidal_th}) causes the IA to undo some of the lensing signal.

Given a value of $A_{\rm IA}$ and following Eq. (\ref{eq:tidal_th}), we couple the tidal field $s_{ij, 2D}$ with the complex intrinsic ellipticities ${\boldsymbol \epsilon}^{\rm IA}$, from which we   compute observed ellipticities as: 
\begin{eqnarray}
{\boldsymbol \epsilon}^{\rm obs} = \frac{{\boldsymbol \epsilon}^{\rm int} + {\boldsymbol g}}{1 + {\boldsymbol \epsilon}^{\rm int, *}{\boldsymbol g}} \,, {\rm with \, \, } 
{\boldsymbol \epsilon}^{\rm int}  = \frac{{\boldsymbol \epsilon}^{\rm IA} + {\boldsymbol \epsilon}^{\rm ran}}{1 + {\boldsymbol \epsilon}^{\rm IA, *}{\boldsymbol \epsilon^{\rm ran}}}.
\label{eq:eps_obs}
\end{eqnarray}
In the above expressions, the denominators guarantee that no ellipticity component exceeds unity.
The complex spin-2 reduced shear ${\boldsymbol g} \equiv (\gamma_1 + {\rm i} \gamma_2)/(1+\kappa)$ is computed from the shear  $(\gamma_{1/2}$) and convergence ($\kappa$) maps, interpolated at the galaxy positions and redshifts. The complex  spin-2 random orientation term ${\boldsymbol \epsilon}^{\rm ran}$ is drawn from two Gaussians (one per component) with their standard deviations provided by the shape noise level detailed in Table \ref{table:survey}. We further constraint  the random ellipticity to satisfy $|{\boldsymbol \epsilon}^{\rm ran}| \le 1.0$. 

We recall that since our simulated galaxy catalogues trace the dark matter density, our default IA-infusion method  is consistent with the $\delta$-NLA model (see Sec. \ref{subsec:IA_th_ext}). Analytical predictions for the two-point functions are more involved in this case \citep[see][]{Blazek2019} and not yet available on the public {\sc cosmoSIS} release. We therefore validate our methods with the standard NLA first, which we infuse by `correcting' the $\delta$-NLA measurement with Eq. (\ref{eq:tidal_th_deltaNLA}), i.e. by replacing ${\boldsymbol \epsilon}^{\rm IA}$ by ${\boldsymbol \epsilon}^{\rm IA}/(1 + b_{\rm TA} \delta)$ in our simulations. Once again, these are less realistic but better suited for validation of the measured $\gamma$-2PCFs against a theoretical model; after this is established, we use the $\delta-$NLA catalogues as our fiducial infusion model.

To be clear, the current version makes no differentiation between galaxy colours or type, and instead treats the full sample as a single population for which we measure an effective alignment signal, similar to the colour-free incarnation of the NLA model \citep[see][for an example with a red/blue split]{DESY1_IA_Samuroff}.

\subsection{Data vector}
\label{subsec:data_vector}

We first set out to validate our IA-infusion method with a two-point analysis, in which we compare the $\gamma$-2PCFs extracted from the simulations with the analytical NLA predictions described in Eq. (\ref{eq:xipm_th}) and Sec. \ref{subsec:IA_th}.  
The measurements are carried out with the public codes {\sc Treecorr} \citep{treecorr}  and {\sc Athena} \citep{athena}, which both compute, for each tomographic bin combination $(\alpha,\beta)$, the estimator $\widehat{\xi_{\pm}^{\alpha\beta}}(\vartheta)$ as: 
 \begin{eqnarray}
\widehat{\xi_{\pm}^{\alpha\beta}}(\vartheta) = \frac{\sum_{ab} W_a W_b \, \Bigg[ \epsilon_{a,{\rm t}}^{\alpha}({\boldsymbol \theta}_a) \, \epsilon_{b,{\rm t}}^{\beta}({\boldsymbol  \theta}_b)  \pm  \epsilon_{a,\times}^{\alpha}({\boldsymbol \theta}_a) \, \epsilon_{b,\times}^{\beta}({\boldsymbol \theta}_b)  \Bigg]\Delta \vartheta_{ab}}{\sum_{ab} W_a W_b} . 
\label{eq:xipm_data}
\end{eqnarray}
The lensing weights $W_{a/b}$  are set to  $1.0$ throughout this paper, and  $\epsilon_{a,{\rm t/\times}}^{\alpha}({\boldsymbol \theta}_a)$ refers to the tangential/cross component of the observed ellipticity from galaxy `$a$' with respect to the centre of the line that connects galaxies $a$ and $b$. The binning operator $\Delta \vartheta_{ab}$ is set to unity if the angular distance  between the galaxy pair  falls inside the $\vartheta$ bin, and to zero otherwise; the sums run over all galaxy pairs in the selected tomographic samples. The measurements on the KiDS-like samples are carried with {\sc Treecorr} in 9 logarithmically-spaced $\vartheta$-bins using min/max separation angles of 0.5 and 475.5 arcmin,  and a bin-slope parameter set to 0.01. In contrast, the {\it Euclid}-like measurements are performed with {\sc Athena}  in 10 logarithmically-spaced $\vartheta$-bins spanning the range 
$[0.1 - 300]'$, with a tree opening angle set to the {\sc Athena} recommended value of  1.7 degrees. 

We also explore the effect of IA on three non-Gaussian statistics constructed from aperture mass maps $M_{\rm ap}(\boldsymbol \theta)$ \citep{Schneider1996}.
Closely following the measurement methods described in \citet{Martinet20} and \citet{Martinet21}, these maps are constructed directly from the galaxy catalogues by stacking the tangential component of ellipticities around  pixels at position $\boldsymbol \theta$, weighted by an aperture filter $Q(\theta, \theta_{\rm ap}, x_c)$: 
 \begin{eqnarray}
M_{\rm ap}(\boldsymbol \theta) = \frac{1}{n_{\rm gal}(\boldsymbol \theta) \sum_a W_a}\sum_a \epsilon_{a,{\rm t}}({\boldsymbol \theta}, {\boldsymbol \theta}_a) Q(|{\boldsymbol \theta} - {\boldsymbol \theta}_a|, \theta_{\rm ap}, x_c) W_a.
\label{eq:Map}
\end{eqnarray}
$n_{\rm gal}(\boldsymbol \theta) $ is the  galaxy density in the filter centred at $\boldsymbol \theta$, $\boldsymbol \theta_a$ is the position of galaxy $a$, and the tangential ellipticity with respect to the aperture centre  is computed as $\epsilon_{a,{\rm t}}({\boldsymbol \theta}, {\boldsymbol \theta}_a)= -[\epsilon_1({\boldsymbol \theta}_a)\ {\rm cos}(2\phi({\boldsymbol \theta}, {\boldsymbol \theta}_a))+\epsilon_2({\boldsymbol \theta}_a)\ {\rm sin}(2\phi({\boldsymbol \theta}, {\boldsymbol \theta}_a))]$,  where $\phi({\boldsymbol \theta}, {\boldsymbol \theta}_a)$ is the angle between both coordinates.
We choose the same aperture filter as the previous two references:
\begin{eqnarray}
Q(x) = \frac{{\rm tanh}(x/x_c)}{x/x_c} \big[1 + {\rm exp}(6 - 150x) + {\rm exp}(- 47 + 50x)\big]^{-1} \,,
\label{eq:Q}
\end{eqnarray}
 which is optimised for detecting haloes following an NFW profile \citep{Schirmer2007}. Here again, $x=\theta/\theta_{\rm ap}$, $\theta$ is the distance to the filter centre,
 $x_c = 0.15$ and $\theta_{\rm ap}$ is set to $12.5'$  for Stage III mocks, and to $10'$ for Stage IV mocks, as this parameter is also optimized against shape noise.

The local variance in these maps is evaluated from the magnitude of the ellipticities: 
\begin{eqnarray}
\sigma^2_{\rm ap}(\boldsymbol \theta) = \frac{1}{2 n_{\rm gal}^2(\boldsymbol \theta) (\sum_a W_a)^2} \sum_a |{\boldsymbol \epsilon}_{a}|^2 Q^2(|{\boldsymbol \theta} - {\boldsymbol \theta}_a|,  \theta_{\rm ap}, x_c) W_a^2\,, 
\label{eq:MapNoise}
\end{eqnarray}
allowing us to construct signal-to-noise maps $\mathcal{S}/\mathcal{N}(\boldsymbol \theta)\equiv M_{\rm ap}(\boldsymbol \theta) / \sigma_{\rm ap}(\boldsymbol \theta)$ in which we identify and count, in bins of \snr:
\begin{itemize}
    \item{peaks ($N_{\rm peaks}$), defined as pixels with  \snr values larger than their 8 neighbours,}
    \item{pixel values ($N_{\rm pix}$), in other words  the full PDF\footnote{These PDF are sometimes referred to the `lensing PDF' or the 1D distribution in the literature.},}
    \item{minima ($N_{\rm min}$), defined as pixels with \snr values smaller than their 8 neighbours.}
\end{itemize}
These make up the three aperture map-based (non-Gaussian) data vectors investigated in this paper; we focus our attention on the peak statistics at first and come back to minima and PDF later on. Closely following \citet{Martinet20}, these measurements are conducted on each of the five tomographic bins, but also on joint catalogues constructed from the union between multiple bins, i.e. 1$\cup$2, 1$\cup$3... 4$\cup$5,  since these `cross-bins' contain a significant amount of additional  information that is not captured by the `single-bins tomography' configuration (referred to as the `auto-bins'). For the {\it Euclid}-like measurements we additionally include the union of more than two bins, e.g. 1$\cup$2$\cup$3, 1$\cup$2$\cup$4 ... 1$\cup$2$\cup$3$\cup$4$\cup$5.
Finally, it is shown in \citet{Martinet20} that Eqs. (\ref{eq:Map}) and (\ref{eq:MapNoise}) can be evaluated with fast convolutions on ellipticity grids as opposed to performing the  sum over the exact galaxy positions in the aperture. This comes at a negligible cost in accuracy, and we exploit this speed-up approach for the analysis of our {\it Euclid}-like catalogues. Note that in this case we directly re-use the measurements from the SLICS and the {\it cosmo}-SLICS simulations presented in \citet{Martinet20} to estimate the covariance matrix and model the cosmology dependence, which by design have the same survey properties as the Stage-IV IA-infused mocks introduced in this paper.

\section{Results: Validation with $\gamma$-2PCFs}
\label{sec:validation}

\begin{figure*}
\begin{center}
\includegraphics[width=8.2in]{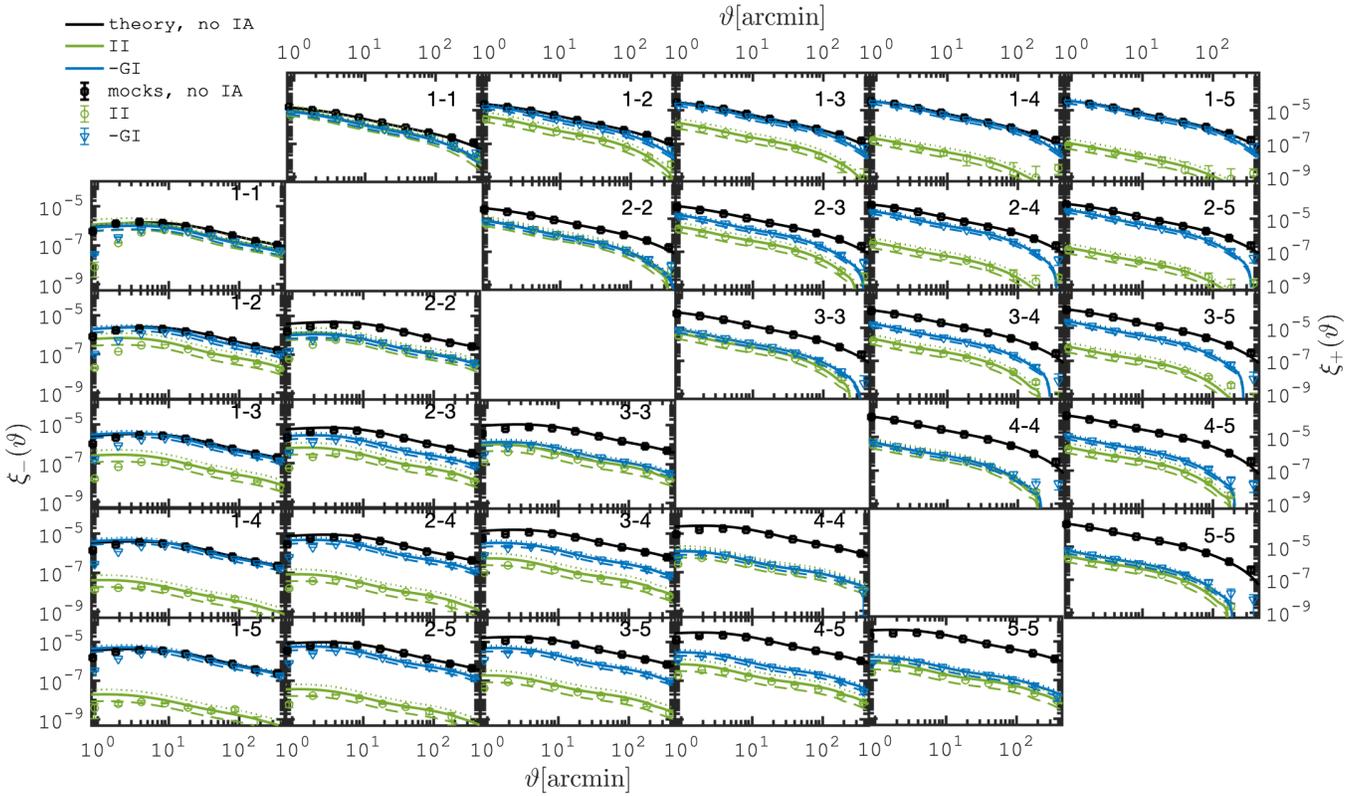}
\caption{Different components of the $\gamma$-2PCFs $\xi_+^{ij}$ ({\it upper right panels}) and $\xi_-^{ij}$ ({\it lower left}), for different tomographic bin combinations $(i,j)$.  The black squares show the measurements  from 50 noise-free light-cones without IA, while the blue triangles and green circles show the $-GI$ and $II$ terms measured from the IA-infused simulations, respectively. The $A_{\rm IA}$ parameter is set to 1.5 in the simulations, while the dashed, solid and dotted line show the NLA predictions for  $A_{\rm IA}=1.0$, $1.5$ and $2.0$, respectively. The tidal fields are smoothed with a Gaussian kernel with a smoothing scale of $\sigma_{\rm G}=0.1 h^{-1}$Mpc. In this figure, the error bars indicate the error on the mean;  these are difficult to distinguish here given their small sizes, but become more apparent later on when presenting ratios.}
\label{fig:xi}
\end{center}
\end{figure*}

\begin{figure*}
\begin{center}
\includegraphics[width=3.5in]{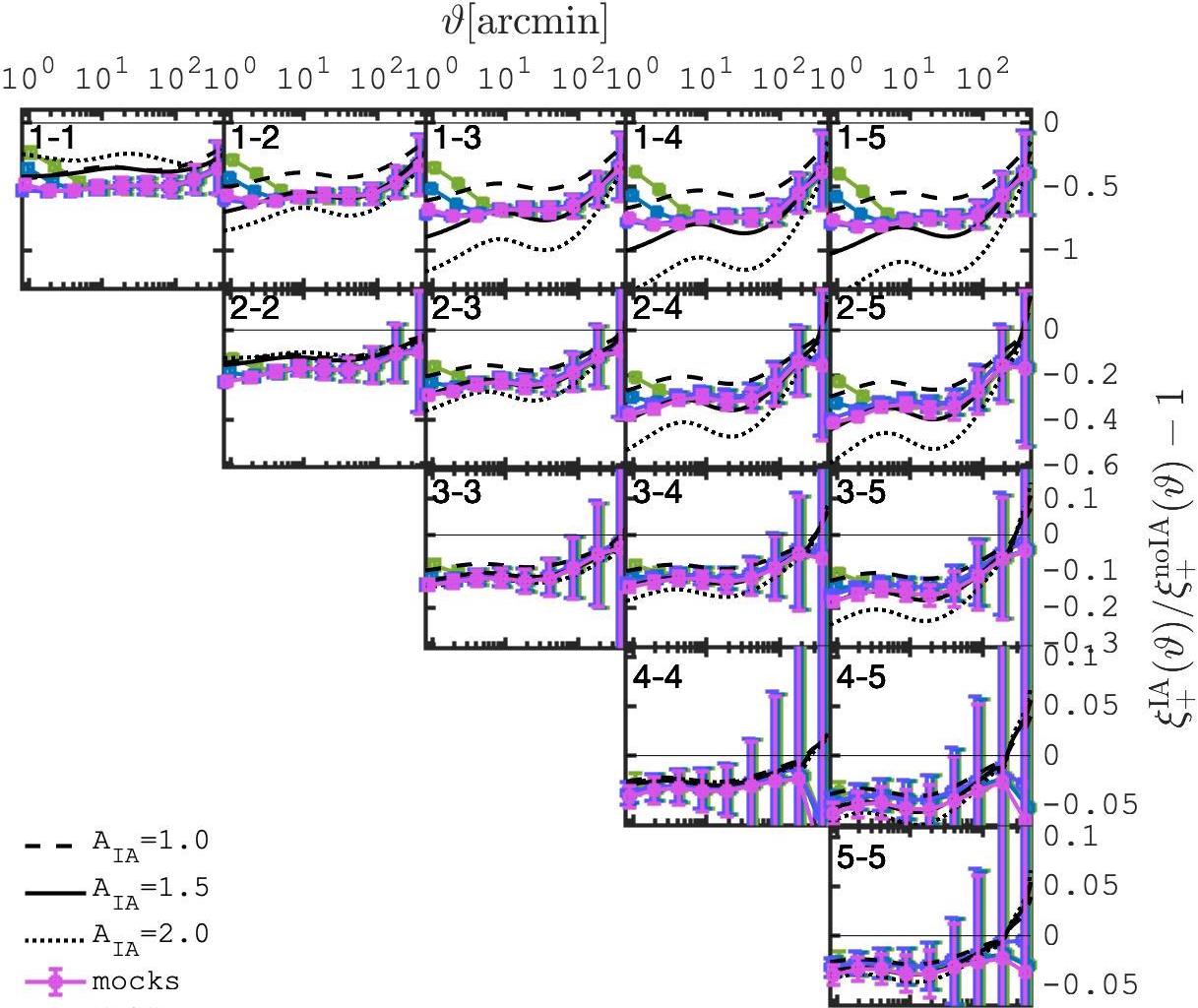}
\includegraphics[width=3.5in]{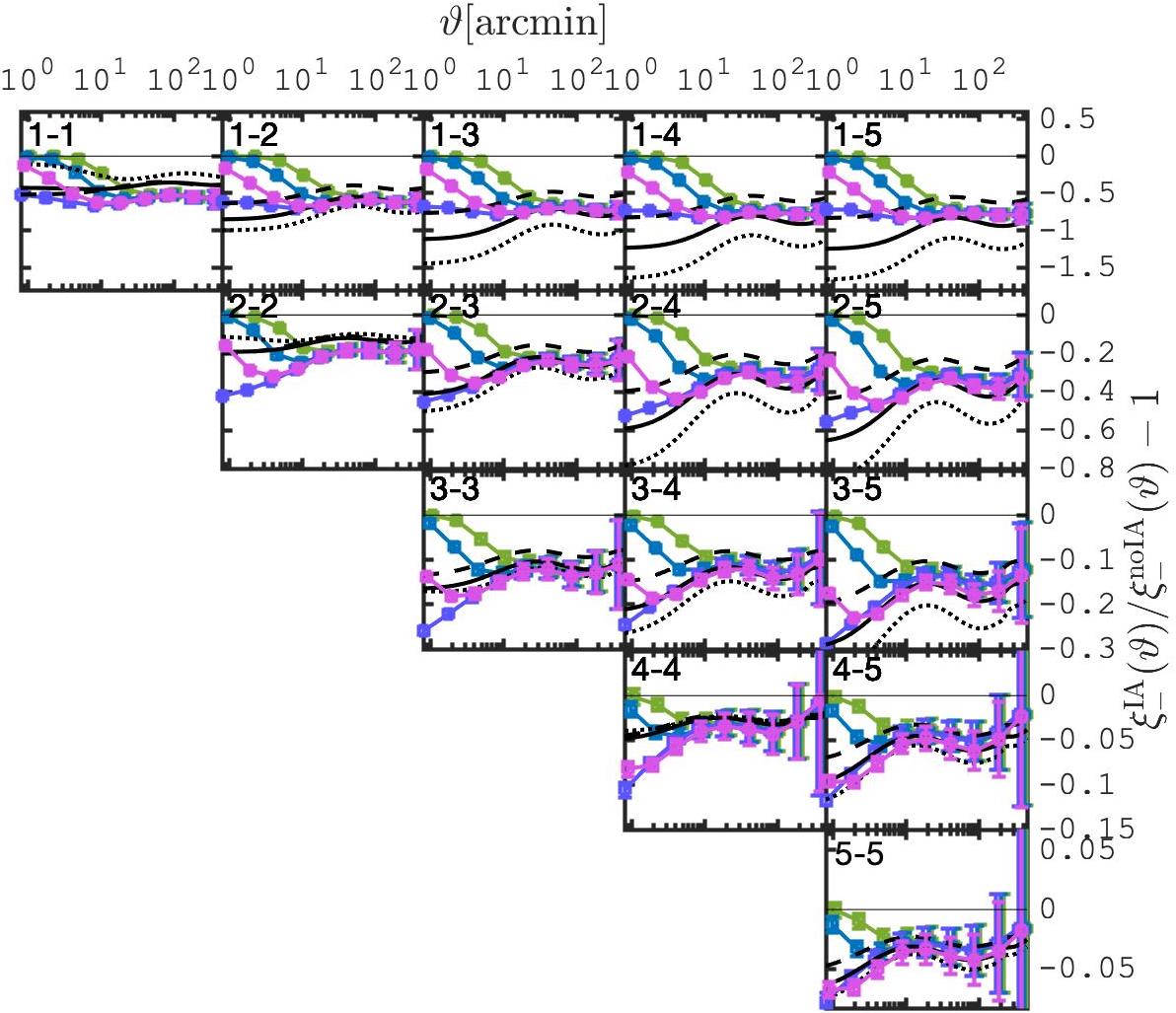}
\caption{Fractional difference between the shear 2PCFs with and without the IA signal, $\xi_{\pm}^{\rm IA}/\xi_{\pm}^{\rm noIA}-1$, in the NLA model. The theory lines show the predictions with three values of the $A_{\rm IA}$ parameter indicated in the legend, while the measurements are for $A_{\rm IA}$=1.5 with different smoothing scales: the magenta squares are our default model, with $\sigma_{\rm G}$=$0.1 h^{-1}$Mpc, while the other coloured curves show $\sigma_{\rm G}$ = 0.5 (green), 0.25 (blue) and 0.0$h^{-1}$Mpc (purple). The error bars show the error on the mean measured from 50 light-cones.  Note the different scaling of the $y$-axis. As for Fig. \ref{fig:xi}, no shape noise is included in this figure, to better exhibit the IA signature.  The thin horizontal line represents the $A_{\rm IA}$=0.0 case.}
\label{fig:xi_frac}
\end{center}
\end{figure*}

\begin{figure*}
\begin{center}
\includegraphics[width=3.3in]{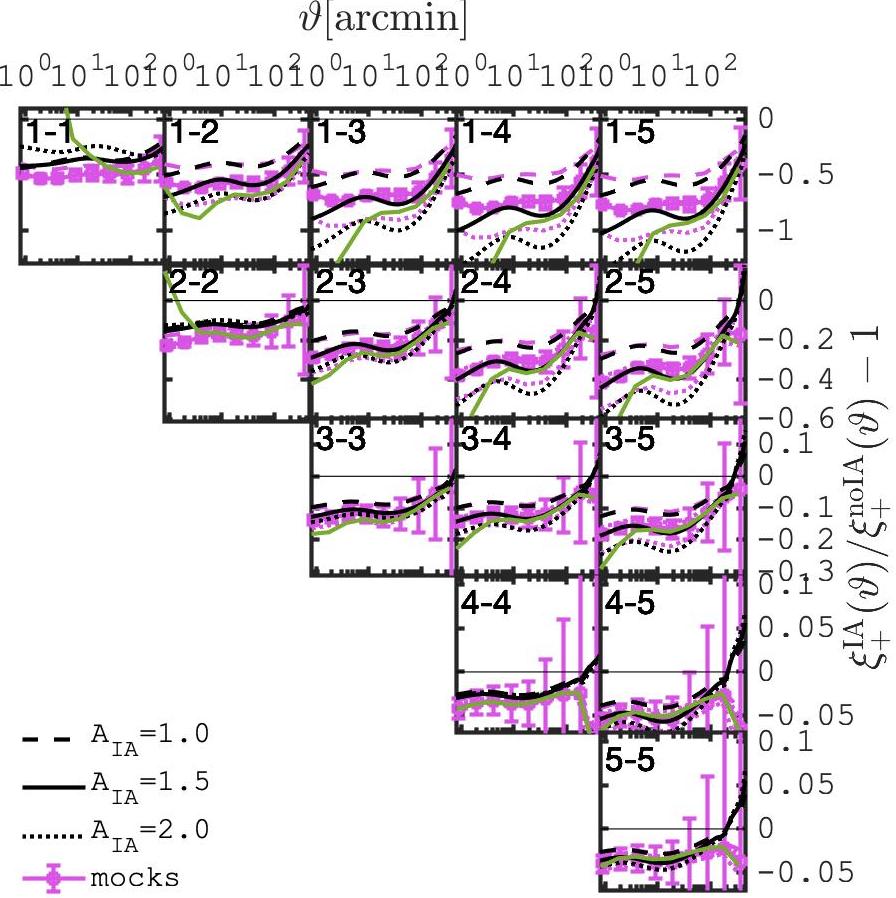}
\includegraphics[width=3.3in]{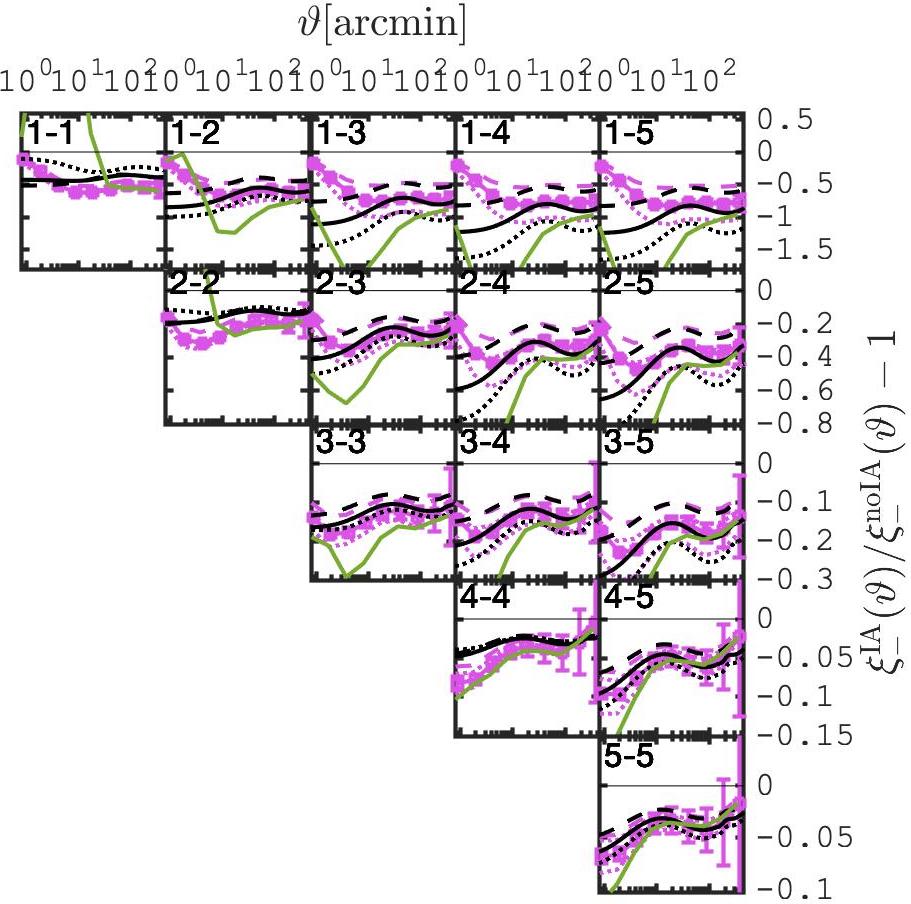}
\caption{Similar to Fig. \ref{fig:xi_frac}, but here the strength of the $A_{\rm IA}$ parameter  is varied in the mocks,  taking values of $2.0$ (magenta dotted lines), $1.5$ (magenta symbols with solid lines) and $1.0$ (magenta dashed lines) -- the same as those assumed for the three NLA theory lines (shown in black).  The smoothing scale is fixed to  $\sigma_{\rm G}$=$0.1 h^{-1}$Mpc. Again, no shape noise is included here, and the error is about the mean, measured from 50 light-cones. The green line shows the results from the infused $\delta$-NLA model.}
\label{fig:xi_frac_varAIA}
\end{center}
\end{figure*}

This section presents a comparison between  the $\gamma$-2PCFs measurements from our NLA-infused KiDS-like catalogues and the analytical predictions. 

\subsection{$GG$, $GI$ \& $II$}
\label{subsec:IA-2PCF}

We show in Fig. \ref{fig:xi} the different components of the $\gamma$-2PCF, namely the $GG$, $GI$ and the $II$ terms, measured from the KiDS-like catalogues. The fiducial infusion model adopted here assumes a smoothing scale of 0.1 $h^{-1}$Mpc, a value that we justify in the next section and which corresponds to a mass of $4.8\times 10^{9}h^{-1}M_\odot$ at redshift zero. The shape noise is switched off in this measurement, easing the comparison with the theoretical model. As shown in \citet[][see fig. 6 therein]{cosmoSLICS}, the $GG$ signal (black squares) matches the theoretical predictions to a few percent at most angular scales, with some power loss at small physical scales due to limitations in the mass resolutions of the $N$-body runs.  This is mostly visible in the $\xi_-$ panel for $\vartheta<3$ arcmin, and hence these scales should be re-calibrated or excluded when entering a cosmological inference based on real data. The $GI$ and $II$ terms (blue triangles and green circles, respectively) also show a similar overall agreement, undershooting the theory line only for $\vartheta<10$ arcmin in $\xi_-$. 

 The $II$ and $GI$ terms, the later of which dominates the contamination  within the LNA and $\delta$-NLA models, are accurately captured\footnote{This statement will need to be revisited for more complex scenarios, since it was shown in \citet{DESY3_Secco} that the $II$ term can be dominant in the TATT model.}.

While the $II$ contribution increases the overall signal, the $GI$ contribution  typically removes power since the intrinsic alignment of galaxies and cosmic shear are anti-correlated. Moreover, the relative importance of the IA contribution to the cosmic shear signal rapidly decreases towards higher redshift, as expected from the NLA model:  the $II$ term becomes weaker there, while at the same time the $GG$ contribution increases by about an order of magnitude. Therefore, the $II$ signal is only important for the lowest redshift bins in auto-correlation, while  the $GI$ contribution is the largest in the cross-redshift bins, when correlating a mix of very high and very low redshift galaxies.
\subsection{$\xi_\pm^{\rm IA}$ vs $\xi_\pm^{\rm no IA}$}
\label{subsubsec:sigma_G}

Fig. \ref{fig:xi_frac} shows the theoretical predictions for the fractional difference between the $\gamma$-2PCFs with and without IA -- the different types of black lines represent the NLA model with  $A_{\rm IA}$ = 1.0, 1.5 and 2.0. We also over-plot in Fig. \ref{fig:xi_frac} the measurements from our simulations assuming $A_{\rm IA}$ = 1.5. The magenta squares show the fiducial smoothing scale, while the results obtained with different  values of $\sigma_{\rm G}$ are presented with the other coloured symbols. There is an excellent agreement with the theory line at large angular separation, with some deviations seen in the highly non-linear scales, notably for $\vartheta<10$ arcmin in $\xi_-$, and $\vartheta<3$ arcmin in $\xi_+$. These are the same scales for which the $II$ and $GI$ were found to be inaccurate in the previous section, due to limits in the mass resolution of the simulations. As $\sigma_{\rm G}$ grows, the impact of IA at small-scale is increasingly erased, causing the measurements to reconnect with the no-IA case as $\vartheta \rightarrow 0.0'$, which is mostly visible in $\xi_-$ for the  $\sigma_{\rm G}$ = 0.5 $h^{-1}$Mpc case (green symbols). The agreement with theory is the best for the case without smoothing (purple lines in Fig. \ref{fig:xi_frac}), however for physical reasons it is unlikely that the tidal fields on scales smaller than 100 kpc could influence the galaxy orientations, which themselves extend over tens of kpc. We have selected $\sigma_{\rm G}$ = 0.1$h^{-1}$Mpc for our fiducial model as it traces well the underlying theory predictions for most scales of interest, a value that could be revisited in the future. There is a visible vertical offset between the simulations and the model for bins 1-1 and 2-2, which is due to a slight under-estimation of the $II$ term in the simulations, as reported above. This is expected to have only a small impact on the cosmological inference, given that the signal is maximal at higher redshifts, and that only two out of the 15 tomographic configurations are affected. The source of this  underestimation is unclear, it is likely due to the large projection length (257.5 $h^{-1}$ Mpc) along the redshift direction involved in the construction of our tidal field maps.  Indeed, we could possibly have obtained a better match had we computed the tidal fields from the full particle distribution instead or finer redshift sampling, however these are not available to us and hence we leave this verification for future work.  

It should also be mentioned that the smoothing prescription is a physical criterion that does not appear explicitly in the theoretical NLA model, aside from optional cuts in $k$-modes that are allowed to contribute to the $P_{II}(k)$ and $P_{GI}(k)$ spectra, as in \citet{Blazek2015}. We do not include such selections here. 

\subsubsection{Varying $A_{\rm IA}$}

The final validation step consists in verifying that our infusion model correctly responds to changes in the coupling strength between the intrinsic galaxy shapes and the tidal fields, as controlled by the $A_{\rm IA}$ parameter. This is tested by consistently modifying the parameter value in the model (Eq. \ref{eq:tidal_th}) and at the infusion stage (Eqs. \ref{eq:Pk_II_th} and \ref{eq:Pk_GI_th}). We present our results for  $A_{\rm IA}$=1.0, 1.5 and 2.0 in Fig. \ref{fig:xi_frac_varAIA}, where we observe that the scaling of the signal with $A_{\rm IA}$ is as expected, with some deviations mostly in the auto-tomographic bins, and a good overall agreement. 
Achieving an accurate $A_{\rm IA}$ scaling is an important milestone: we can hereafter vary this parameter in the simulated lensing catalogues and use Eq. (\ref{eq:eps_obs}) to construct new mock observations, from which we can interpret cosmic shear data with a flexible IA sector, for any weak lensing estimator.  This is therefore compatible with joint [$\gamma$-2PCFs; Peaks] analyses as in \citet{Martinet18} and \cite{HD20}, but now augmented with a new joint $A_{\rm IA}$ marginalisation capacity. 

Of course, this presupposes that the infusion model adequately describes the physical IA at play in the real Universe,  which is still highly uncertain: according to some recent studies \citep[e.g.][]{Blazek2015,Fortuna2020,DESY1_Troxel} the standard NLA model is disfavored over more complex models, whereas \citet{DESY3_Secco} find that although consistent, the TATT model is unnecessarily flexible for the analysis of the DES-Y3 data and in fact degrades the cosmological constraints compared to the NLA case. If required, additional features can be infused in Eq. (\ref{eq:tidal_th}) including redshift scaling \citep[e.g. with the $\eta_{\rm IA}$ parameter, see eq. 15 of][]{KiDS1000_Asgari}, or even completely different physical models that can be constructed from  the tidal tensor such as the TATT.
One of such extensions has been introduced in Sec. \ref{subsec:IA_th_ext}, the $\delta$-NLA model,  and results in significantly different IA signatures at small scales, shown by the green lines in  Fig. \ref{fig:xi_frac_varAIA}. This has been computed by fixing the smoothing length to $0.1h^{-1}$Mpc in both models, a choice that might be revisited in the future, but nevertheless the deviations from the NLA are clear and visible at all scales.

\section{Results: Impact on $M_{\rm ap}$ statistics}
\label{sec:Peaks}

\subsection{Peak count + IA: Impact on Stage-III surveys}
 \label{subsec:PeaksKiDS}
 
\begin{figure}
\begin{center}
\includegraphics[width=3.7in]{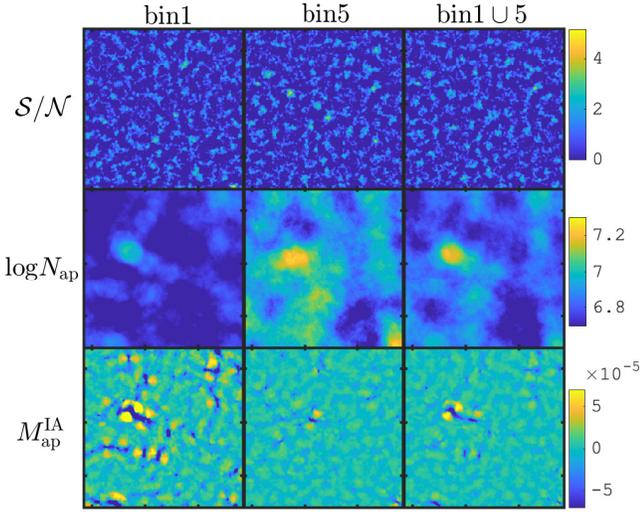}
\caption{({\it upper panels:}) $\mathcal{S}/\mathcal{N}$ maps, constructed from different tomographic bins in one of the KiDS-1000-like simulations,  including intrinsic alignments  in the galaxy orientations computed from the tidal fields shown in Fig. \ref{fig:tidalator}, for a smoothing length of 0.1$h^{-1}$ Mpc and $A_{\rm IA}=1.5$.  
({\it middle panels:}) Aperture number count map, constructed from the same galaxy samples. ({\it bottom panels:}) Contamination from the intrinsic alignment of the galaxies (shown in the middle panels) to the aperture mass maps (shown in the upper panels).
Note that the upper and lower panels do not show the same quantity, whence the differences in the color scale.}
\label{fig:Map}
\end{center}
\end{figure}

\begin{figure}
\begin{center}
\includegraphics[width=3.3in]{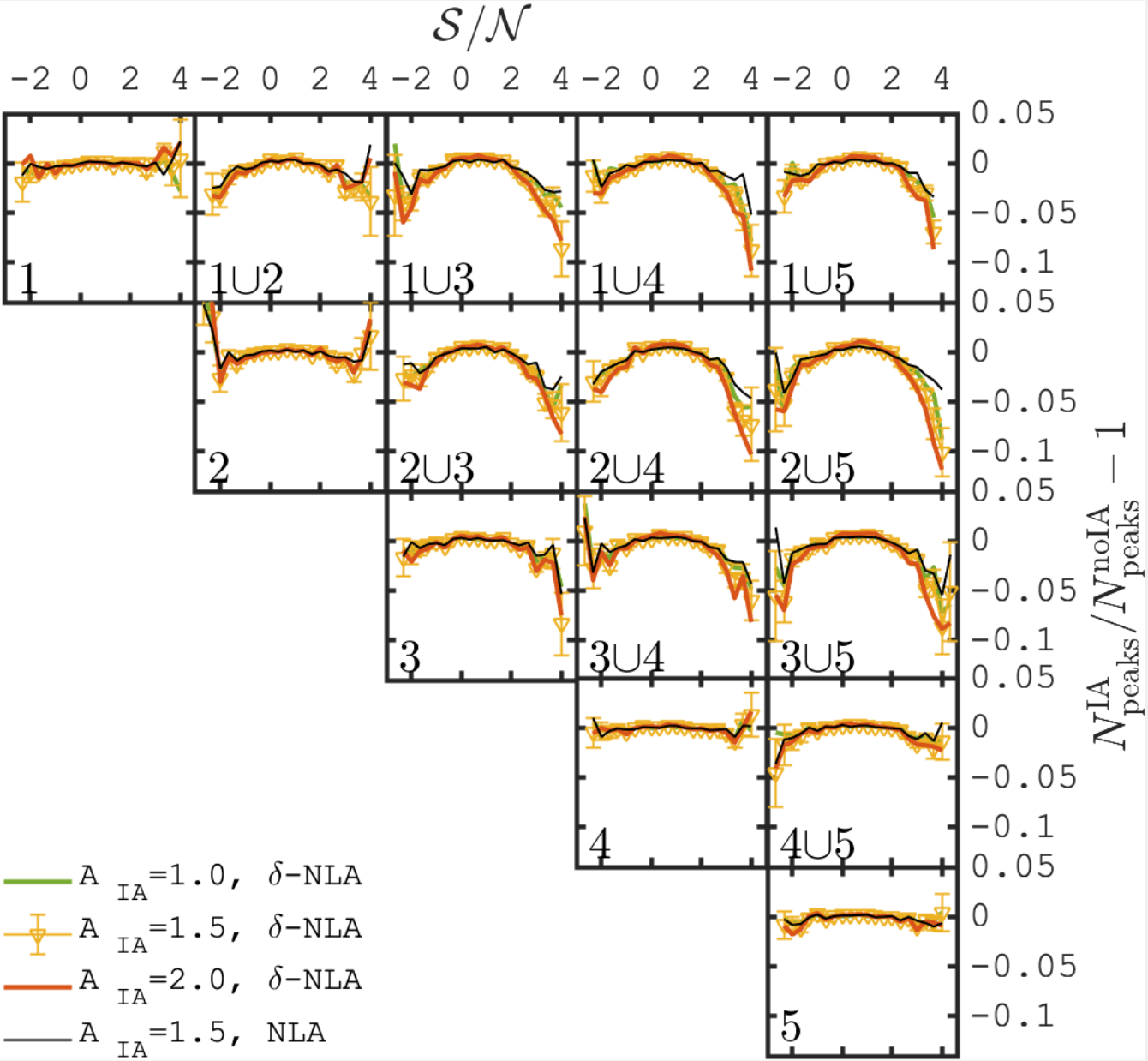}
\caption{Fractional difference between the number of lensing peaks with and without the IA signal, measured from the KiDS-like simulations. Here again, the measurements are obtained after smoothing the tidal field with a smoothing scale of $0.1 h^{-1}$Mpc.  The error bars about the yellow symbols show the error on the mean, measured from 50 light-cones with shape noise, for $A_{\rm IA}=1.0$ (green), $1.5$ (yellow) and $2.0$ (red). }
\label{fig:peaks_frac_KiDS}
\end{center}
\end{figure}

\begin{figure}
\begin{center}
\hspace{9mm}\includegraphics[width=2.9in]{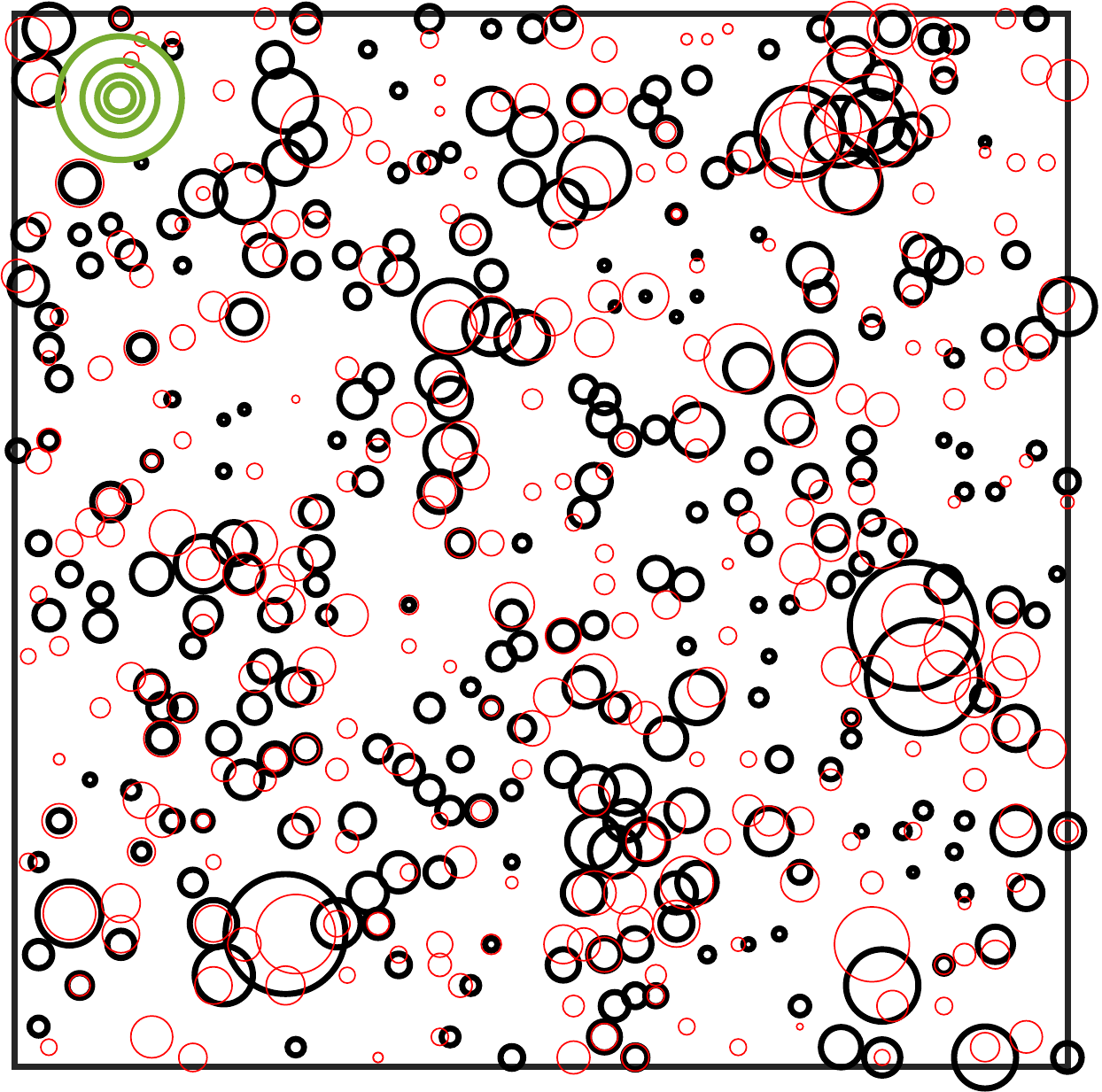}
\includegraphics[width=3.3in]{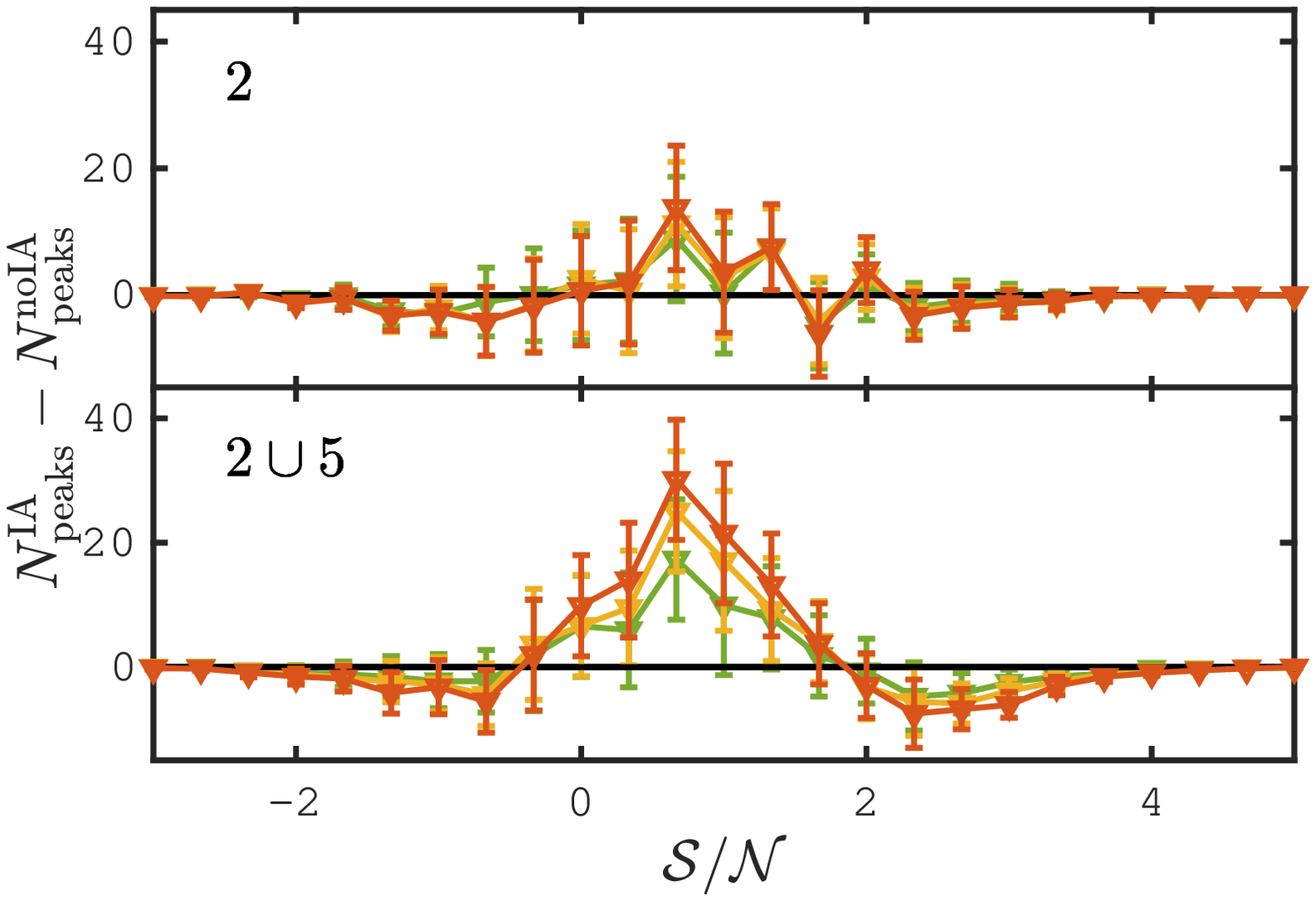}
\caption{({\it upper:}) Position and height of the peaks measured in the tomographic bin combination 2$\cup$5, with $A_{\rm IA}$=2.0 (red) and 0.0 (black), inside a 1.0 deg$^2$ patch. The four concentric circles indicate the sizes for \snr=1, 2, 3 and 4. ({\it lower:}) Difference between the peak function with and without IA as measured from our 50 KiDS-like catalogues, for the tomographic bin $2$ and the combination 2$\cup$5, shown for  $A_{\rm IA}$ = 1.0 (green), 1.5 (yellow) and 2.0 (orange).}
\label{fig:dNpeaks_IA}
\end{center}
\end{figure}

Having validated our IA infusion model with the $\gamma$-2PCFs in the last section, we now turn our investigation towards the aperture mass statistics defined in Sec. \ref{subsec:data_vector}. With the shape noise turned on hereafter,
we construct   \snr maps and measure the lensing peak  function, $N_{\rm peaks}$, in the KiDS-like catalogues. We carry out this measurement for all tomographic bins and their pair-wise combinations, with and without the IA. Following \citet{HD20}, we use \snr bins sizes of 1/3, here over the wider range of \mbox{ [-3$\le$\snr\!\!$\le$4]}.  We consider hereafter the $\delta$-NLA model with $\sigma_{\rm G}=$ 0.1 $h^{-1}$Mpc as our fiducial case, being more realistic than the standard NLA, but discuss some NLA results as well.  The multiple panels in  Fig. \ref{fig:Map} illustrate the different contributions to the observed aperture mass map.  The top panels show the \snr maps from tomographic bins 1, 5 and 1$\cup$5;  the middle panels  map out  the aperture number count $N_{\rm ap}$, obtained by setting the ellipticities to unity in Eq. (\ref{eq:Map}); the bottom panels present the IA contamination map, $M_{\rm ap}^{\rm IA}$, present in the observed \snr shown in the top panels. We can compute $M_{\rm ap}^{\rm IA}$ either from the difference between the measured $M_{\rm ap}$ with and without IA, or by passing the pure IA signal to (Eq. \ref{eq:Map}) instead of the observed ellipticities.   Note the strong correlation between $N_{\rm ap}$ and $M_{\rm ap}^{\rm IA}$, expected from the coupling between the IA ellipticities, the tidal field and the mass over-density. The relative impact of IA on the observed $M_{\rm ap}$ is only a few percent over most of the pixels and is  therefore not visible by eye. As for two-point correlations, the IA contribution is higher in low redshift slices (bin 1 compared to bin 5), which both contributes in the combination of slices (bin 1$\cup$5). The effect on the aperture mass statistics further depends on the noise level, which is the lowest for combined bins. 

When counting peaks with and without the IA however, we notice a stronger effect on the high $\mathcal{S}/\mathcal{N}$ peaks, as shown in Fig. \ref{fig:peaks_frac_KiDS}. This suppression is particularly strong in the cross-tomographic bins,  causing in some cases a 10\% suppression of peaks with $\mathcal{S}/\mathcal{N}$=4.0. This can be understood by the fact that high $\mathcal{S}/\mathcal{N}$ peaks are found in regions of high foreground density, where substantial alignment of galaxies is produced, leading to a large contribution from a $GI$-like term\footnote{The expressions $GG$, $II$ and $GI$ strictly apply only to two-point functions, however we can generalise this concept for other types of measurements. Here, we use the expression $GI$-like to describe the contribution to the total peak count statistics coming from cosmic shear signal ($G$) of high-redshift  galaxies, with the intrinsic signal ($I$) from the low-redshift ones.}.
There is also a noticeable effect on peaks with large negative $\mathcal{S}/\mathcal{N}$ values, which is not surprising given that these are highly correlated with the high $\mathcal{S}/\mathcal{N}$ peaks \citep{Martinet18}. Similarly to the $\gamma$-2PCFs, the cross-redshift bins are more severely affected by IA whereas the auto-tomographic bins are relatively immune; this could prove helpful in a data analysis for choosing elements of the measurement vector that are less affected by uncontrolled IA systematics \citep[as done in][and further explored in Sec. \ref{sec:Peaks}]{HD20}. Although this conclusion certainly depends on the alignment model in place, one would still expect the cross-redshift bins to be most affected because of the $GI$-like terms. For instance, impact from the NLA model is shown by the black lines in Fig. \ref{fig:peaks_frac_KiDS}, which exhibits a milder effect than the $\delta$-NLA model, consistent with the $\gamma$-PCFs. Alternatively, if, for example, a non-perturbative IA model \citep[e.g.][]{Fortuna2020} was to be  considered instead of the current $\delta$-NLA, much larger alignments around halos could occur; the same applies also for higher order operators of the tidal field, however, we leave this investigation for future work.

Changing the smoothing scale from 0.1 to 0.5 $h^{-1}$Mpc has almost no effect of the results presented in Fig. \ref{fig:peaks_frac_KiDS}. This is likely caused by the additional smoothing inherent to aperture map statistics (see Eq. \ref{eq:Map}), which suppresses features smaller than the filter scale. These two smoothing scales correspond to angular scales of less than 2 arcmin (except for at the very lowest redshifts, where no lensing signal originate anyway) and are therefore not well distinguishable in the aperture mass maps. Note that this conclusion will not hold for smaller aperture filter sizes however, for which we expect to recover stronger IA effects for smaller  $\sigma_{\rm G}$.
 
When it comes to understanding what happens to individual peaks in presence of IA,  \citet{Kacprzak2016} suggest that the suppression of large peaks  arises from foreground satellite galaxies in-falling onto massive foreground clusters, but being incorrectly included in the background source galaxy sample. Inspired by \citet{SchneiderBridle2010}, their model assumes that the radial alignment of these cluster-member satellites would tend to undo the tangential cosmic shear signal imparted by the said cluster onto the true background galaxies. This hypothesis is however not supported by the observational analysis of \citet{Sifon+15} who measured negligible radial alignment of satellite galaxies around galaxy clusters. In addition, Fig. \ref{fig:peaks_frac_KiDS} reveals that IA affect the peak function even in absence of  cluster member contamination, since this is not explicitly included in our simulations. To be precise, our method introduces the equivalent of a small local contamination, since all simulated galaxies that populate a given mass sheet receive the same IA contribution. This, however, would not explain the IA effect observed in cross-tomographic bins, which are well separated in redshift; the presence of a $GI$-like term is a more accurate explanation. 

Another difference with the \citet{Kacprzak2016} model is that while the {\it fractional} effect is the largest on large positive and negative  $\mathcal{S}/\mathcal{N}$ peaks, we find that the {\it absolute} effect is larger for peaks with $\mathcal{S}/\mathcal{N}\sim [0.0-1.0]$, as seen in Fig. \ref{fig:dNpeaks_IA}. The bottom panel of this figure shows the difference $N_{\rm peaks}^{\rm IA}-N_{\rm peaks}^{\rm noIA}$ instead of the ratio, for the auto-tomographic bin 2 and the combination 2$\cup$5. The impact is negligible in the former but strong in the latter, and demonstrates that IA tend to increase the number of peaks in the range $-0.5<$ \snr$<2$ and suppress those outside that range. The `bubble plot' in the upper panel shows the peak positions found in the 2$\cup$5 catalogue with (red)  and without  IA (black), where the size of each bubble scales with the peak's \snr value. We observe a clear correlation between the large red and black circles, indicating that both cases identify common foreground over-densities with high significance. However the largest black circles generally have their red counterparts slightly smaller in size, corresponding to a reduction in \snr observed in the peak function.   Moreover their positions are slightly shifted, but not as much as for the smaller peaks, for which there are many red circles with no black counterpart and vice versa. Because of their increased number, the low-\snr peaks are in an absolute sense  maximally affected by IA, as already seen in the lower panel of Fig.~\ref{fig:dNpeaks_IA}, but relatively speaking, these only affect a small fraction of the peaks.
We further note that although this detailed inspection  is presented here only for the bin combination 2$\cup$5, others cross-bins  show similar but milder features, while in the auto-tomographic cases  the relative effect of IA is consistent with noise, as shown in the diagonal panels of Fig. \ref{fig:peaks_frac_KiDS}. 

We can conclude from this investigation that the impact of IA is non-negligible in the peak count analyses of Stage-III lensing survey as it suppresses systematically the high-$\mathcal{S}/\mathcal{N}$ end of data vector, which is also highly sensitive to cosmology \citep{Martinet18}. With their cluster-member contamination model,  \citet{Kacprzak2016} reports a small 1.5\% shift in $S_8$ due to IA in a non-tomographic setting, which was also adopted in the KiDS-450 peak count analysis of  \citet{Martinet18}.  
The tomographic DES-Y1 analysis of \citet{HD20} modelled the IA with a halo occupation distribution model, aligning the central galaxies with the shape of the host dark matter haloes, without any mis-alignment scatter. They find a sub-percent impact on $S_8$, but their model includes IA only in the bins 1, 2 and 1$\cup$2 (out of four bins), albeit with a stronger relative effect on the peak function. In light of the IA infusion model investigated in this paper, it is clear that all cross-tomographic bins are affected by the IA, which suggests that the three analyses  above-mentioned could have under-estimated the impact of IA. The two earlier analyses were carried in a  non-tomographic setting and therefore are as vulnerable to IA contamination, as we demonstrate in the next section.

Comparing the amplitude of the impact with $\gamma$-2PCFs, we expect that completely neglecting the IA in a peak count case is less catastrophic: whereas the IA affects $\xi_{\pm}$ at all angular bins by sometimes 50-100\% (Fig. \ref{fig:xi_frac_varAIA}), only the largest peaks are affected here, and by no more than 5-10\% (Fig. \ref{fig:peaks_frac_KiDS}). We show in the next section how this affects the inferred cosmological constraints for the upcoming generation of lensing surveys.

\subsection{Peak count + IA: Impact on Stage-IV surveys}
 \label{subsec:PeaksEuclid}

\begin{figure}
\begin{center}
\includegraphics[width=2.7in]{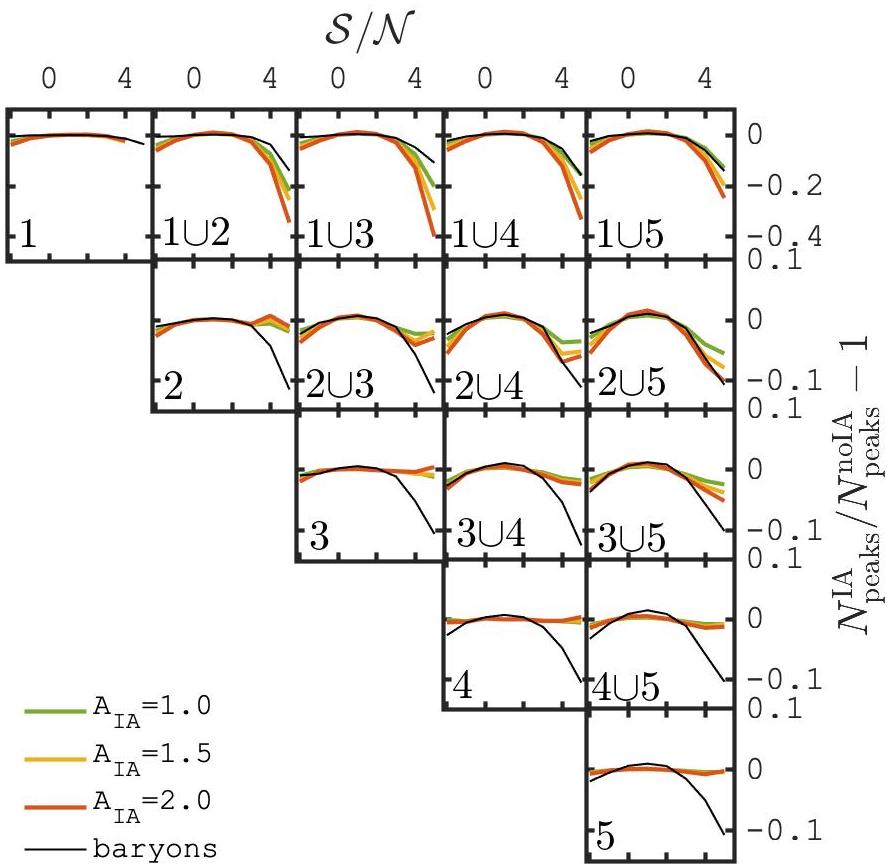}
\includegraphics[width=2.7in]{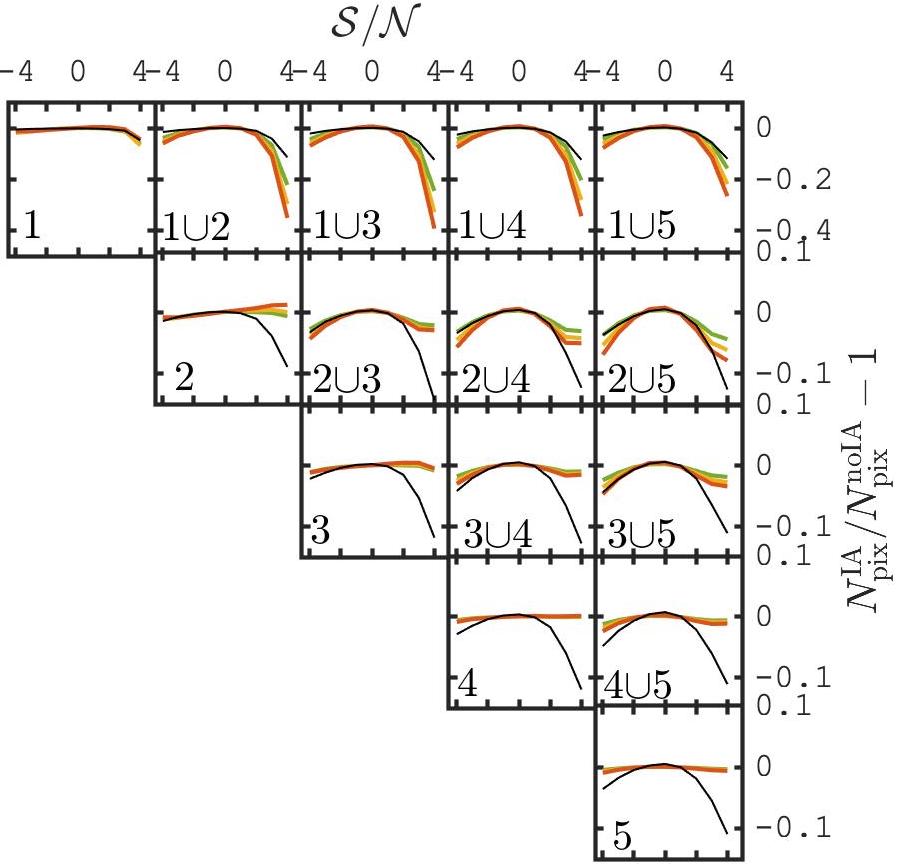}
\includegraphics[width=2.7in]{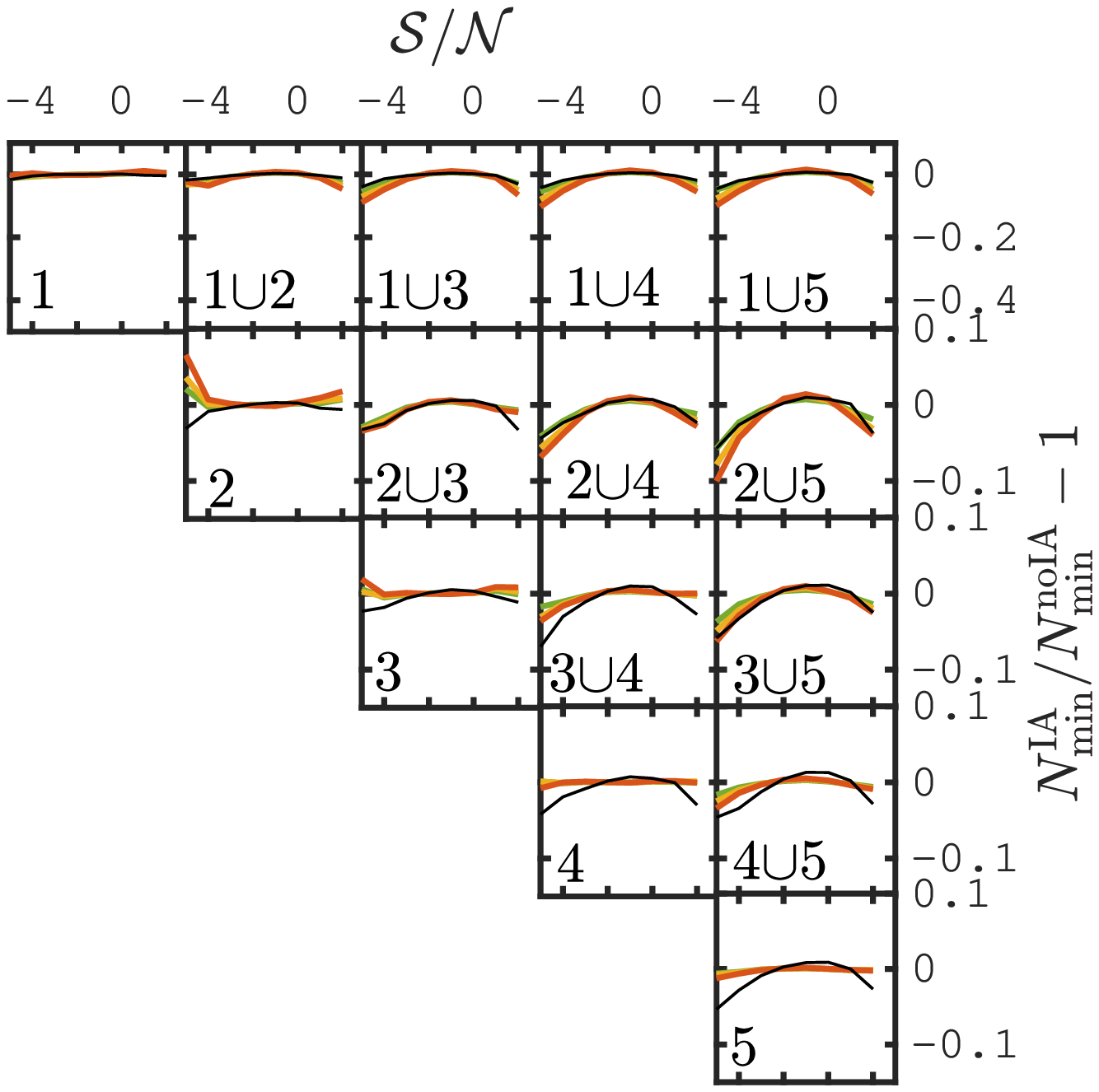}
\caption{({\it upper:}) Same as Fig. \ref{fig:peaks_frac_KiDS}, but measured from the 10 {\it Euclid}-like simulations, each averaged over 5 noise realisations. The thin black lines indicate the impact from the baryonic feedback reported in \citet{Martinet21}, which generally affects the signal more severely than the IA except at the lowest redshifts. The different colored lines show the effect of varying $A_{\rm IA}$. ({\it middle and lower:}) Same as upper panel, but for lensing PDF ($N_{\rm pix}$) and minima ($N_{\rm min}$), respectively. A  smoothing of 0.1$h^{-1}$Mpc is used here, however these curves are mostly unchanged if employing $\sigma_{\rm G}$=0.5$h^{-1}$Mpc instead.  Note the change in the $y$-axis scaling for the lowest redshift bin.}
\label{fig:peaks_frac_Euclid}
\end{center}
\end{figure}

\begin{figure*}
\begin{center}
\includegraphics[width=3.5in]{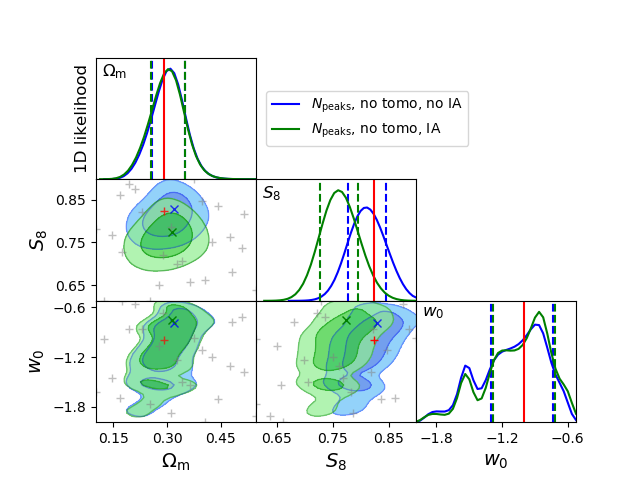}
\includegraphics[width=3.5in]{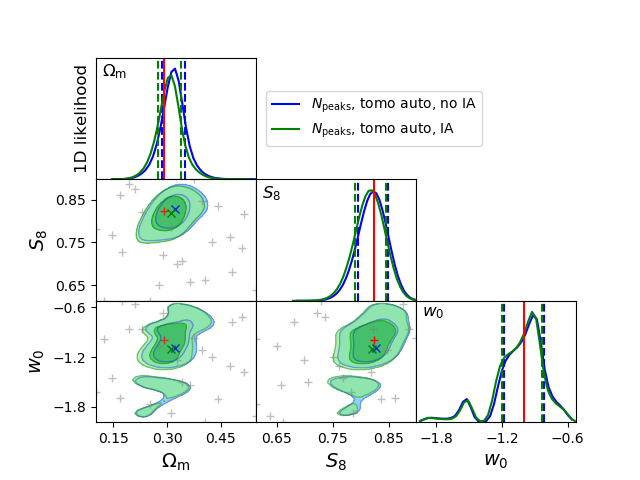}
\includegraphics[width=3.5in]{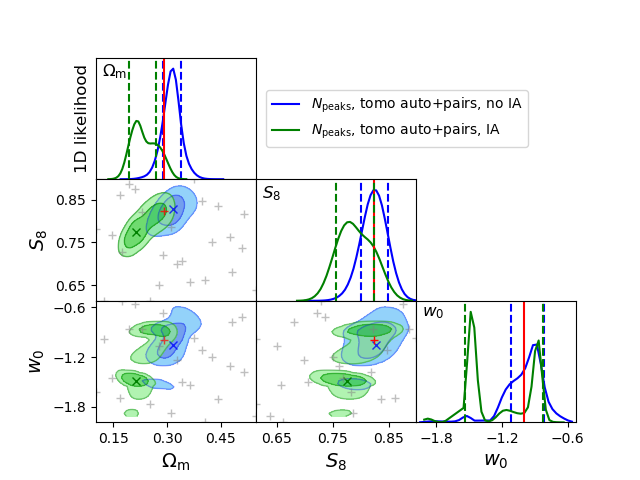}
\includegraphics[width=3.5in]{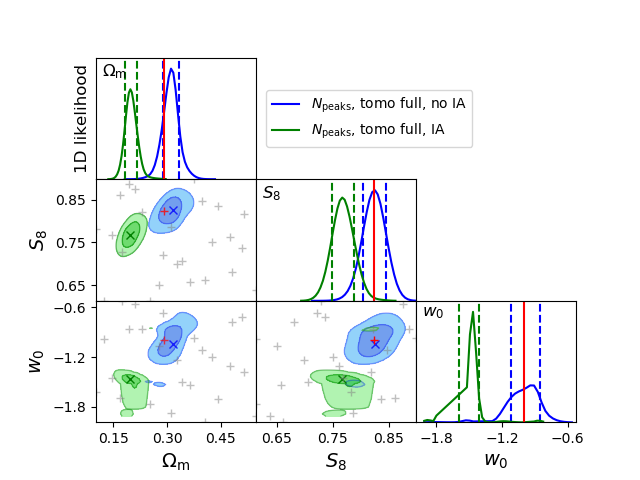}
\includegraphics[width=3.5in]{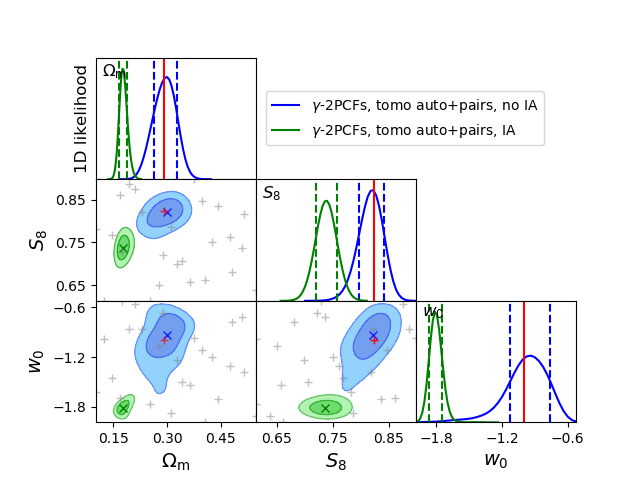}
\includegraphics[width=3.5in]{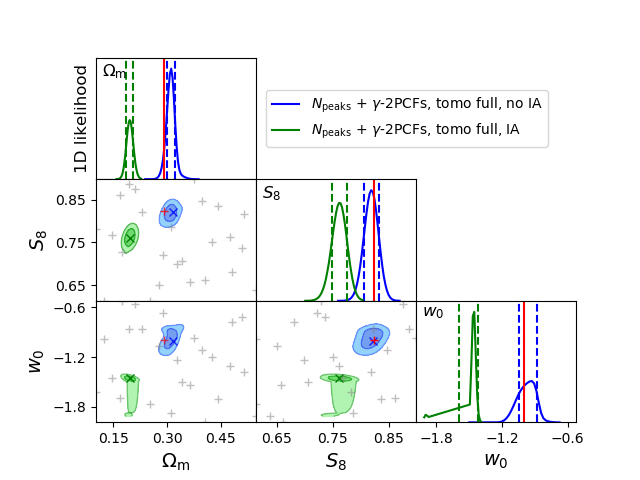}
\caption{Bias on the $w$CDM  cosmological parameters from analysing 100 sq. deg. of  IA-infused {\it Euclid}-like data with models calibrated on simulations without IA.
The upper four panels are for peak without tomography  ({\it upper left}), with auto-tomographic bins ({\it upper right}), with auto-tomographic bins + bin pairs ({\it middle left}) and with all bin combinations  ({\it middle right}). 
The full tomographic 2PCFs analysis ({\it lower left}) and joint analysis ({\it lower right}) are significantly more affected by IA in comparison, especially  in $w_0$. The red lines and `$+$' indicate the input cosmology while the grey `$+$' symbols indicate the {\it cosmo}-SLICS training nodes.}
\label{fig:like}
\end{center}
\end{figure*}
 
 \begin{table}
   \centering
   \caption{Different survey configurations investigated with Stage-IV cosmology inference pipeline. For each of these, we consider three aperture mass statistics $N_\alpha$(\snr\!\!), namely peaks ($\alpha$=peaks), minima ($\alpha$=min) and lensing PDF ($\alpha$=pix).}
   \tabcolsep=0.11cm
      \begin{tabular}{@{} ccccccccc @{}} 
      \hline
      \hline
              & \multicolumn{4}{c}{$N_\alpha$(\snr)} & \multicolumn{3}{c}{$\gamma$-2PCFs} &  \\
     case &  no-tomo & auto & pairs & all &  no-tomo& auto & cross    & $N_{\rm d}$ \\
     \hline
     (i)  &$\surd$& & & & &  & & 8 \\
     (ii)  && $\surd$& & & & & & 40 \\
     (iii)  && $\surd$& $\surd$&  & & & & 120 \\
     (iv)  && $\surd$&$\surd$ &$\surd$ &&& & 200 \\
     (v)  && & & & $\surd$& &  & 16\\
     (vi)  && & & & & $\surd$ & &80\\
     (vii)  && & & & &$\surd$ &$\surd$ &240\\
     (viii)  &$\surd$& & & & $\surd$ & & & 24\\
     (ix)  & &$\surd$& $\surd$& & &$\surd$ &$\surd$ & 360\\
     (x)  & & $\surd$&$\surd$& $\surd$& & $\surd$ & $\surd$ &440\\
    \hline 
    \hline
    \end{tabular}
    \label{table:configuration}
\end{table}

\begin{table*} 
\caption[]{Bias on the inferred cosmology caused by the intrinsic alignments of galaxies, if left unaccounted for, for the different cases listed in Table \ref{table:configuration}. Parameter shifts are reported as `$\Delta$', and can be contrasted with the precision expected for a 100 deg$^2$ of {\it Euclid}-like data, reported as $\delta\equiv\sigma_{\rm stat}$. Numbers in parenthesis show  the shifts $\Delta$ in units of statistical errors $\delta$, and the percentage of the marginalised error with respect to the input parameter value. An $A_{\rm IA}=1.5$ and a smoothing scale of 0.1$h^{-1}$Mpc are assumed here. The cases labelled as `M20' are extracted from table 1 in \citet{Martinet20}.}
\centering 
\begin{tabular}{clcccccc} 
\hline 
case& estimator(s)& $\Delta S_8$ & $\Delta w_0$ & $\Delta \Omega_{\rm m}$ & $\delta S_8$ & $\delta w_0$ & $\delta \Omega_{\rm m}$ \\ 
\hline 
\hline
 (i)&$N_{\rm peaks}$, no tomo & -0.052 (-1.4) & 0.038  (0.1) & -0.007 (-0.1) & 0.035 (4.2\%) & 0.284 (28.4\%) & 0.047 (16.2\%)\\
 (ii)&$N_{\rm peaks}$, tomo auto &-0.003 (-0.1) & -0.009  (-0.0)& -0.01  (-0.3)& 0.027 (3.3\%) & 0.186 (18.6\%) & 0.033 (11.3\%)  \\
 (iii)&$N_{\rm peaks}$, tomo auto+pairs &  -0.042 (-1.2)& -0.606 (-1.7) & -0.103 (-2.7) & 0.034 (4.1\%) & 0.358 (35.8\%) & 0.038 (13.3\%)\\
 (iv)&$N_{\rm peaks}$, tomo full & -0.058  (-2.9) & -0.524 (-5.9) & -0.113 (-6.6) & 0.02 (2.5\%) & 0.089 (8.9\%) & 0.017 (5.7\%)\\
 (M20)&$N_{\rm peaks}$, tomo full, no IA & -- & -- & -- &0.022 (2.7\%) & 0.130 (13\%) & 0.024 (8\%) \\
\hline
 (v) &$\gamma$-2PCFs, no tomo & -0.042 (-1.0) & 0.000 (0.0)& -0.069 (-1.1) & 0.042 (5.1\%) & 0.307 (30.7\%) & 0.062 (21.2\%) \\
 (vi)&$\gamma$-2PCFs, tomo auto &  -0.006 (-0.2) & -0.037 (-0.1) & 0.005  (0.1)& 0.026 (3.2\%) & 0.253 (25.3\%) & 0.044 (15.1\%)  \\
 (vii)&$\gamma$-2PCFs, tomo auto+pairs & -0.083  (-4.4) & -0.878 (-15.1) & -0.124 (-11.6) & 0.019 (2.3\%) & 0.058 (5.8\%) & 0.011 (3.9\%)\\
 (M20)&$\gamma$-2PCFs, tomo auto+pairs, no IA  & -- & -- & -- & 0.023 (2.8\%) & 0.190 (19\%) & 0.034 (12\%)\\
\hline
 (viii)&joint, no tomo &  -0.087 (-3.2) & -0.423 (-1.4) & -0.029 (0.6) & 0.027 (3.2\%) & 0.294 (29.4\%) & 0.046 (15.8\%)\\
 (ix)&joint, tomo auto+pairs & -0.067 (-4.5) & -0.553 (-5.2) & -0.118 (-9.8) & 0.015 (1.9\%) & 0.106 (10.6\%) & 0.012 (4.0\%)\\
 (x)&joint, tomo full & -0.057  (-4.1) & -0.527 (-6.0)& -0.116 (-11.6) & 0.014  (1.7\%) & 0.088 (8.8\%) & 0.010 (3.3\%)  \\
 (M20)&joint, tomo full, no IA &    -- & --  & -- & 0.015 (1.8\%)& 0.090 (9\%) & 0.017 (6\%)\\
\hline 
\hline
 (i)&$N_{\rm min}$, no tomo  & -0.051 (-1.3)  & 0.000 (0.0) & 0.000 (0.0) & 0.040 (4.8\%) & 0.273 (27.3\%) & 0.092 (31.7\%) \\
 (ii)&$N_{\rm min}$, tomo auto & 0.000 (0.0) & 0.000  (0.0)& 0.012 (0.2)& 0.029 (3.6\%) & 0.195 (19.5\%) & 0.049 (16.8\%) \\
 (iii)&$N_{\rm min}$, tomo auto+pairs & -0.041 (-1.3) & -0.582 (-1.7) & -0.086 (-2.2) & 0.032 (3.9\%) & 0.347 (34.7\%) & 0.04 (13.7\%)\\
 (iv)&$N_{\rm min}$, tomo full & -0.056  (-2.5) & -0.578 (-10.7) & -0.099 (-3.8) & 0.022 (2.7\%) & 0.054 (5.4\%) & 0.026 (9.1\%) \\
 (M20)&$N_{\rm min}$, tomo full, no IA  & -- & -- & -- & 0.024 (2.9\%)& 0.120 (12\%) & 0.027 (9\%)\\
\hline
 (viii)&joint, no tomo  & -0.088 (-3.1) & -0.037  (-0.2)& -0.024 (-0.5) & 0.028 (3.4\%) & 0.233 (23.3\%) & 0.052 (17.8\%)\\
 (ix)&joint, tomo auto+pairs & -0.069 (-3.8) & -0.506 (-7.1) & -0.129 (-10.8)& 0.018 (2.2\%) & 0.071 (7.1\%) & 0.012 (4.3\%)\\
 (x)&joint, tomo full & -0.079  (-5.6) & -0.528  (-6.0)& -0.127 (-11.5) & 0.014 (1.7\%) & 0.088 (8.8\%) & 0.011 (3.7\%)\\
 (M20)&joint, tomo full, no IA & -- & -- & -- & 0.014 (1.7\%)&0.080 (8\%)&0.017 (6\%)\\
\hline
\hline
 (i)&$N_{\rm pix}$, no tomo  & -0.063 (-2.1) & 0.000 (0.0) & 0.000 (0.0) & 0.030 (3.6\%) & 0.139 (13.9\%) & 0.035 (12.1\%)\\
 (ii)&$N_{\rm pix}$, tomo auto & -0.000  (0.0) & 0.036 (0.2) & -0.015  (-0.5)& 0.026 (3.2\%) & 0.217 (21.7\%) & 0.028 (9.8\%) \\
 (iii)&$N_{\rm pix}$, tomo autopairs & 0.016 (0.6) & 0.039 (0.5) & -0.056 (-1.75) & 0.029 (3.6\%) & 0.075 (7.5\%) & 0.032 (11.2\%)\\
 (iv)&$N_{\rm pix}$, tomo full &0.008 (0.4) & 0.052 (1.9)& -0.069 (-3.1) & 0.021 (2.5\%) & 0.028 (2.8\%) & 0.022 (7.5\%) \\
 (M20)&$N_{\rm pix}$, tomo full, no IA & -- & -- & -- & 0.021 (2.6\%)&0.100 (10\%)&0.019 (7\%) \\
\hline
 (viii)&joint, no tomo  &  -0.081  (-3.1) & 0.028 (0.23) & -0.003 (-0.1) & 0.026 (3.2\%) & 0.118 (11.8\%) & 0.031 (10.7\%) \\
 (ix)&joint, tomo auto+pairs &  -0.060 (-4.3) & -0.596 (-3.7) & -0.093 (-8.5) & 0.014 (1.6\%) & 0.162 (16.2\%) & 0.011 (3.7\%)  \\
 (x)&joint, tomo full & -0.037 (-3.4) & -0.570  (-6.0) & -0.119 (-13.2) & 0.011 (1.3\%) & 0.095 (9.5\%) & 0.009 (2.9\%)\\
 (M20)&joint, tomo full, no IA & -- & -- &  -- & 0.013 (1.5\%)&0.060 (6\%)&0.015 (5\%)\\
\hline
\hline

\label{table:cosmo_bias_01}
\end{tabular} 
\end{table*}

With highly improved statistics, Stage-IV lensing surveys will be increasingly sensitive to systematics and secondary signals. In this section we aim to forecast the impact of IA on a cosmic shear measurement based on 100 square degrees of {\it Euclid}-like data, analysed with the peak count statistics, $\gamma$-2PCFs, and in the joint case. As detailed in Sec. \ref{subsec:WL_cats}, we use the {\it cosmo}-SLICS catalogues to model the cosmology dependence of the signal and the 928 SLICS catalogues to estimate the joint covariance matrix.  The likelihood is modelled as a multivariate $t$-distribution:
\begin{eqnarray}
\mathcal{L}(\boldsymbol{\pi}|\boldsymbol{d}) \propto   \frac{N_{\rm sim} }{2}  {\rm ln}\bigg[1 + \chi^2 / (N_{\rm sim} - 1)\bigg],\mbox{\hspace{5mm} with} 
\label{eq:like}
\end{eqnarray}
\begin{eqnarray}
\chi^2 = \sum [\boldsymbol{x}(\boldsymbol{\pi}) - \boldsymbol{d}]^{\rm T}{\rm Cov}^{-1} [\boldsymbol{x}(\boldsymbol{\pi}) - \boldsymbol{d}] ,\nonumber
\end{eqnarray}
where  $\boldsymbol{d}$  and $\boldsymbol{x}(\boldsymbol{\pi})$ are respectively the measurement and model vectors, ${\boldsymbol{\pi}}$ is the cosmological point at which the model is evaluated, and $N_{\rm sim}$=928.
As explained in \citet{SellentinHeavens}, this choice of  likelihood accurately captures the residual noise caused by the inversion of a covariance matrix estimated from a finite number of independent simulations\footnote{The $t$-distribution likelihood reduces to a multi-variate Gaussian likelihood in the limit of infinite number of simulations.}. With at least hundreds of peaks in each of our $\mathcal{S}/\mathcal{N}$ bins, adopting this  form is justified. The likelihood is sampled within a parameter range that is determined by the {\it cosmo}-SLICS training set:
\begin{eqnarray}
0.10 \le  &\Omega_{\rm m}& \le  0.55\nonumber \\
0.60 \le  &S_8&  \le 0.90\nonumber \\
-2.0 \le  &w_0& \le -0.5\nonumber \\
0.60 \le  &h& \le 0.82
\end{eqnarray}
and is evaluated by interpolating the measurements at the {\it cosmo}-SLICS nodes with radial basis functions. We follow the approach developed for hydrodynamical simulations in \citet{Martinet21} and infuse the bias measured in the IA catalogues to the measurement data vector. This ensures that differences between the mocks used for the $w$CDM model and those used for IA infusion (for instance the different sampling of galaxy positions) are not biasing our model, nor our estimate of the resulting bias on cosmology. The interpolation is performed in a two-step approach on a regular grid of 40 points for each parameter, the second step refining the parameter step to the hyperspace where the likelihood is non-zero.

The relative effect of IA on peak statistics is measured from  our {\it Euclid}-like simulations in 8 bins spanning the range [-2.5$\le$\snr\!\!$\le$5.5], and shown in the upper panel of Fig. \ref{fig:peaks_frac_Euclid}. The trends seen in the KiDS-like data are similar and amplified here, with elements in the cross-terms bins affected by up to 40\% (see e.g. the bin $1\cup$3).   Once again, IA are the strongest at low redshift and in cross-tomographic bins. Over-plotted on these measurements are the effect of baryonic feedback measured from the Magneticum simulations in \citet{Martinet21}, which unlike IA is approximately constant for all redshift bins. Since the cosmological information extracted by lensing is the highest at high redshift, we could expect peak count statistics from deep surveys to be more affected by baryonic feedback than IA, however this conclusion could of course vary when considering models different from the  $\delta$-NLA and Magneticum.

Fig. \ref{fig:peaks_frac_Euclid} also presents the results for lensing PDF ($N_{\rm pix}$) and minima ($N_{\rm min}$) measured from the same simulations and \snr$\!\!(\boldsymbol{\theta}$) maps, with the exception that the former uses 9 bins in the range [-4$\le$\snr\!\!$\le$5], and the latter 8 bins with [-5$\le$\snr\!\!$\le$3], as in \citet{Martinet20}. While the lensing PDF responds to IA in a manner almost identical to the peaks, the lensing minima seems to be more immune, with at most a 10\% effect even in the lowest redshift bins. We could therefore expect that lensing minima would incur a smaller cosmological bias from unmodelled IA, since these are the strongest in high-density regions, which are mostly avoided by this statistics. We show next that although this reasoning has some logic to it, it does not always hold in practice.

Combining the data vectors, the $w$CDM models and the covariance matrices, we now forecast Stage-IV parameter constraints and compare the marginalised {\it a posteriori} distributions after running the likelihood pipeline on data, with and without the alignments.
Any observed difference can be interpreted as a bias on the inferred cosmology caused by the inclusion of IA. We explore different analysis strategies including aperture mass statistics and $\gamma$-2PCFs in a variety of tomographic choices, for the purpose of identifying the configurations that are least affected. As in \citet{Martinet20}, we reject the two largest $\vartheta$-bins in $\xi_+$ to protect us from residual finite simulation box effects,  and further remove the two smallest $\vartheta$-bins in $\xi_-$ as they are subject to uncertain non-linear physics. For each aperture mass statistics $N_\alpha$(\snr\!\!), with $\alpha$= `peaks', `min' or `pix', we consider four distinct cases summarised in Table \ref{table:configuration} (cases i-iv). We further compare these results to the $\gamma$-2PCFs with cases (v-vii) , and the joint probes [$\gamma$-2PCFs; $N_\alpha$(\snr\!\!)] in cases (viii-x).

The results for cases (i), (ii), (iii), (iv) and (vii) with the peak count statistics are shown in Fig. \ref{fig:like}. As first found in \citet{Martinet20}, tomographic peak count analyses can improve by a factor of three the precision on $w_0$, and by a factor of 2.0 that of $S_8$, in absence of systematics. This can be seen by comparing the size of the marginalised posterior distribution  in the different panels of Fig. \ref{fig:like}, more precisely the unbiased cases, and from comparing the marginal constraints reported in Table \ref{table:cosmo_bias_01} in the `no IA' cases.  We expect that the inclusion of additional systematics (e.g. photometric redshifts, baryons) will further deteriorate both $\gamma$-2PCFs and peaks constraints, but this can be achieved without biasing the inferred value if modelled correctly.

We draw the attention on the difference between the biased (green contours) and the unbiased (blue contours) cases, which illustrate the impact IA could have on the inferred cosmology if left unmodelled, assuming $A_{\rm IA}$ = 1.5. There is a clear shift towards lower $S_8$ and $\Omega_{\rm m}$ values in most cases, as reported  in Table \ref{table:cosmo_bias_01}. However, these are mild for peak statistics, in comparison with the bias experienced by the $\gamma$-2PCFs (-1.0\% and -6.3\%, vs. -8,8\% and -31.1\%, for $S_8$ and $\Omega_{\rm m}$ respectively). It is encouraging to see that except for the full tomographic case, the dark energy parameter $w_0$ is less affected by the IA. Whether this secondary spike observed in the full tomography is a real IA - $w_0$ degeneracy or noise caused by our relatively coarse cosmological sampling  remains to be demonstrated in future work based on the next generation of $w$CDM weak lensing simulations, however it is consistent with the shift seen in the $\gamma$-2PCFs (case vii). \citet{Blazek2019} finds similar results, i.e. that unmodelled IA tends to lower $S_8$ by as much as $\Delta S_8=0.1$, and the dark energy equation of state by $\Delta w_0=0.5$. Note that their setup also include a redshift evolution, namely $w(a) = w_0 + w_a(1-a)$, and $w_a$ is shifted by 7$\sigma$ to -3.0, compared to the input value of 0.0. 

We note from Fig. \ref{fig:like} that none of the projected posteriors are prior dominated, however the residual noise observed in the shape is caused by the sparse cosmological sampling and could be improved with more {\it cosmo}-SLICS nodes.

The results for all cases and all estimators are also summarised in Table \ref{table:cosmo_bias_01} (see Appendix \ref{sec:2PCF_IA} for an equivalent table measured from different $A_{\rm IA}$ values).  In the full tomographic joint [$\gamma$-2PCFs;$N_{\rm peaks}$] analysis explored in this paper (case x), an unmodelled $\delta$-NLA IA signal would bias $S_8$ by $\sim$7\%,   $\Omega_{\rm m}$ by $\sim$40\%, and $w_0$ by $\sim$53\%, driven by both the $\gamma$-2PCFs and peaks. In the joint  [$\gamma$-2PCFs;$N_{\rm min}$] analysis, $S_8$,  $\Omega_{\rm m}$ and $w_0$ could be biased by 9.6\%, 43.7\% and 52.8\%, respectively, and similar results are obtained from the [$\gamma$-2PCFs;$N_{\rm pix}$] case (with 4.5\%, 41.1\% and 57.0\%). Contrary to our expectations, lensing minima alone show a similar bias in $S_8$ compared to peaks, but the lensing PDF alone are relatively immune, with a shift of less than 1\%. Most non-Gaussian statistics show a secondary peak develop in $w_0$, which can bias the measurement beyond 50\%.  The effect on $\Omega_{\rm m}$ are the smallest for the lensing PDF, reaching  23.8\% (compared to 34.1\% for minima and 38.9\% for peaks).  As the reported biases correspond to the shift in the maximum of the likelihood between the IA and no IA cases, part of the bias could be due to the sparsity of the cosmological parameter space, although it is dense enough to get the right order of magnitude. We also observe that the inclusion of IA affects the size of the marginalised errors, as the best fit values lie in a different region in parameter space, therefore changing the relative importance of signal and shape noise. 

Comparing now the bias ($\Delta$) to the precision ($\delta$) on the parameters in Table \ref{table:cosmo_bias_01}, we note that the shifts are ranging from the 0-15$\sigma$, the worst cases being the error on $w_0$ in the joint full tomographic settings, where the precision is the highest. We notice that the posterior distribution about this parameter are in this case rather noisy, developing a secondary peak, which inflates significantly this shift. Nevertheless, all parameters are significantly affected by IA, with levels that vary with the choice of estimator and tomographic bin, and no case is fully protected.  
The average biases on $S_8$, $w_0$ and $\Omega_{\rm m}$ are $2.2\sigma_{\rm stat}$, $2.8\sigma_{\rm stat}$ and $4.2\sigma_{\rm stat}$, the latter being systematically severely affected. 

Note that scaling these results to a full {\it Euclid} area of 15,000 deg$^2$ would improve the precision `$\delta$' by a factor of $\sqrt{15,000/100} \sim 12.2$ compared to the current estimates, while leaving the IA-induced bias unchanged. In that case, a 0.3$\sigma$ shift would become a 3.6$\sigma$ bias, which already is catastrophic for the inference. 

To be highlighted, the auto-tomographic analysis of the lensing minima is mostly immune to the IA, with no shift recorded in $S_8$ nor $w_0$, and a 4\% shift in $\Omega_{\rm m}$. Another robust case is the auto-tomographic peaks case, where $\Delta S_8$ is about a tenth of $\delta S_8$, $\Delta w_0\sim 0.05\sigma$, but $\Delta \Omega_{\rm m}\sim0.3\sigma$ already. These measurements are already competitive with the $\gamma$-2PCFs projected constraints, but nowhere near the combined-probe results, especially that of [$N_{\rm pix}$; $\gamma$-2PCFs], which can reach precision of 0.06 on $w_0$, 0.013 on $S_8$ and 0.015 on $\Omega_{\rm m}$, but on which unmodeled IA would shift the results by 6.0, 3.4 and 13.2$\sigma$, respectively.  In light of these, it is clear that IA forward  modelling will be critical for these alternative statistics.

When comparing the global impact of an unmodelled $\delta$-NLA IA signal versus the baryons feedback presented in \citet[][see their table 2]{Martinet20}, we notice that IA affects the $\gamma$-2PCF more severely for all parameters, with  ($\Delta S_8$, $\Delta w_0$, $\Delta \Omega_{\rm m}$) = (-0.083, -0.878, -0.124) for IA compared to (-0.034, 0.037, -0.035) found from analysing the Magneticum simulation with an identical survey setup. Indeed, IA affects all angular scales in $\xi_\pm(\vartheta)$, with an important redshift modulation, causing a large shift in all parameters. The story is similar for all three aperture mass statistics, where we observe that in the full tomographic case, baryons have a smaller impact on $S_8$ (e.g. $\Delta S_8$ = -0.024 vs. -0.058 for IA, as measured with peak statistics), on $\Delta w_0$ (0.035 vs. -0.524), and on $\Delta \Omega_{\rm m}$  (-0.005 vs. -0.113). As for the $\gamma$-2PCFs, this can be explained by the fact that while baryon feedback has a mild redshift dependence, IA exhibits a more complex structure across the different  tomographic bin combinations that is partly  degenerate with the growth of structure, directly affected by both $w_0$ and $\Omega_{\rm m}$ but less so by $S_8$. This picture is also supported by the fact that we find similar biases for baryons and IA in the non-tomographic cases.

We repeated the measurement of these biases when changing  $\sigma_{\rm G}$ from 0.1 to 0.5 $h^{-1}$Mpc  and found that the biases are reduced on average for $\Delta S_8$ and $\Delta w_0$, while  $\Delta \Omega_{\rm m}$ are slightly larger, compared to our fiducial case. The residual noise in the observed marginalised likelihood prevents us from deriving a meaningful scaling relation between the cosmological biases and the smoothing length,  and hence future analyses will likely need to include this as a free parameter in the IA model.

 \subsection{Discussion}
 \label{subsec:discussion}
The methods presented in this paper offer an efficient and accurate avenue to model and mitigate over the uncertain effect and strength of the IA, at the simulation catalogue level.  Weak lensing analyses based on statistics beyond two point functions, which are generally relying on simulation-based inference methods, can now be augmented with a series of dedicated IA-infused mocks from which the secondary signal can be studied for different IA model (NLA, $\delta$-NLA, TATT) as a function of the corresponding parameter values ($A_{\rm IA}$, $b_{\rm TA}, ...$). 

From this point onward, the simplest approach would be to cut out the data elements that are causing the bias in cosmology. For example, the likelihood analysis presented in the last section could be repeated for different selections of \snr bin and choice of tomographic setup, until the bias $\Delta$ falls well under the precision $\delta$ for every cosmological parameters we wish to measure.
This is similar in spirit with the approach adopted by the DES-Y1 cosmic shear analysis of \citet{DESY1_Troxel}, where conservative\footnote{The $\xi_+$ and $\xi_-$ data vectors in \citet{DESY1_Troxel} are truncated for $\vartheta<[4-6]'$ and $<[40-90]'$, respectively.} small-scale cuts are applied such as to guarantee that the effect of baryons and non-linear physics had a controlled and minor impact on the inferred cosmology. This is not adequate for $\gamma$-2PCFs here since most elements are affected by IA, but could be useful for the aperture mass statistics investigated in this paper, for which the effect is localised onto a small number of \snr bins. 

The elements of the data vector mostly contaminated by IA are also highly sensitive to cosmology, since most of the contamination arises from the cross-correlation of the signal of interest and the systematic via the $GI$-like contributions. It is thus more optimal to attempt some form of modelling in the future instead. There exists a few options 
to achieve this. First, one could measure the shift in cosmology and in error bars caused by different realistic IA models separately from the likelihood sampling (e.g. using the results from Table \ref{table:cosmo_bias_01}, then correct the inferred cosmology and inflate error bars accordingly. This approach was adopted by \citet{Martinet18} for the treatment of baryon feedback on peak statistics in the analysis of the KiDS-450 data. The risk here is that there might be interactions between the IA effect and any additional parameters sampled in the full likelihood analysis but excluded from the calculation of Table \ref{table:cosmo_bias_01} (e.g. shape calibration, redshift uncertainty, etc.). 

To avoid this, it is better to model the IA at the level of the measurement data vector and apply a correction factor to the data, as done in e.g. \citet{Kacprzak2016} and in  \citet{HD20}. Repeating for different values of $A_{\rm IA}$ would allow us to estimate the impact on the cosmology inference and the associated systematic uncertainty, to be combined with the statistical error. The main drawback from this method is that the correction factor is computed at a fixed cosmology. This effectively assumes a weak cosmological dependence of the IA signal relative to the underlying signal, an assumption that has yet to be tested and  that could also lead to important biases.

It would be preferable to infuse IA in a series of mocks that cover the full parameter space volume inside of which the likelihood is estimated. This forward modelling approach is more computationally demanding as it requires the calculation of projected tidal fields at every cosmological training node, however the gain in accuracy is guaranteed since it correctly captures the full cosmological dependence of the secondary signal. This is equivalent to the treatment of IA in the $\gamma$-2PCFs pipeline, for which the model (see Eqs. \ref{eq:Pk_II_th} and \ref{eq:Pk_GI_th}) are re-evaluated alongside the $GG$ signal (Eq. \ref{eq:xipm_th}) from the matter power spectrum at each step of the likelihood sampling. This will result in a slight increased uncertainty on the cosmological parameters of interest, as found in  \citet{Zuercher2020a} with an alternative IA-infusion method.

In fact, a number of results presented in this paper can also be obtained from the techniques introduced in \citet{Fluri2019} and \citet{Zuercher2020a}, which is also based on projected mass sheets. The key improvements over their method is that we work at the level of galaxy catalogues, as opposed to lensing maps, and are therefore closer to the data produced by the lensing surveys. Each galaxy is assigned a random orientation, a cosmic shear signal, and an intrinsic alignment extracted from a flexible tidal field-based model \citep[only the NLA model is implemented in][]{Zuercher2020a}. Therefore, combined with the convergence and the local density fields, our method can construct reduced shear values -- the lensing observables -- with more complex IA models, for example by including higher order operators in the tidal fields.

One of the key open questions is whether the accuracy of our infusion method is high enough for the Stage-IV lensing surveys. As presented in \citet{Blazek2019}, next generation data analyses will require flexibility and accuracy in the modelling of the IA sector, without which the inferred cosmological parameters could incur catastrophic biases. In particular, they have demonstrated that an LSST-like dataset infused with the TATT IA model but analysed with the NLA would result in $S_8$ and $w_0$ values that are significantly too low. Adding a redshift dependence to the NLA model reduces these biases down to $\sim 1\sigma$ shifts, however only a full TATT model produces unbiased results. The methods presented in this paper are designed to enable this flexibility, however there are residual discrepancies with the input model that will need to be improved upon, as shown in Fig. \ref{fig:xi_frac}. It is not entirely clear at this stage what causes this (we suspect a projection effect), nor how these deviations will impact the inferred cosmology after marginalisation. For this a full likelihood analysis will need to be carried out. This calls for  further tests to be done in the near future, notably on the convergence of the IA signal as a function of the projection thickness.

We present in this paper our implementation of the NLA and $\delta$-NLA models, however additional physics is likely to be required to analyse upcoming data, in particular the tidal torquing term and a red/blue split dependence. Whereas  the former is within reach with the {\it cosmo}-SLICS simulations, the later could be achieved by applying our methods on external light-cone simulations where these galaxy properties exist, for example in the Millenium\footnote{mpa-garching.mpg.de/galform/virgo/millennium}, MICE\footnote{maia.ice.cat/mice}, {\it Euclid} Flagship \citep{Flagship} or the LSST DC2\footnote{portal.nersc.gov/project/lsst/cosmoDC2/\_README.html} simulations. One could construct different samples based on the colour information (or any other existing galaxy property that could correlate with the IA) and assign an intrinsic alignment per galaxy from projected tidal field, with a coupling that is allowed to vary for different samples. This hybrid scheme would allow for more detailed IA models that hopefully better describe the physics observed in spectroscopic data, but without the need to resolve the halo shapes. This is a particularly powerful advantage, since measuring halo shapes through the inertia matrix can only be achieved for halos resolved with hundreds of particles, introducing a low-mass cut on the halos for which galaxies have a reliable IA assigned. 

We note that having access to the halo information has an additional advantage, in that one could study separately the impact of IA on peaks that correspond to collapsed regions and those that are due to line-of-sight projection, as in \citet{2002ApJ...575..640W}. Additionally, our IA-infusion method can be combined with other IA mitigation techniques, including exploiting the correlation length \citep{Catelan_IA_Tidal}, nulling \citep{PhysRevD.72.043002,2010A&A...517A...4J}, self-calibration \citep{Bernstein_2004, Bernstein_2009, 2020MNRAS.495.3900Y, Sanchez2021}, and using galaxy colour information \citep{2018arXiv180501240T}. So far these techniques have been only applied to the standard $\gamma$-2PCFs and it would be essential to study how they perform in non-Gaussian statistics. In fact, this whole approach could be phrased as a search for the (non-Gaussian) lensing estimator that is the best compromise between capturing a maximum of cosmological information, while being minimally sensitive to a suite of IA models.

Although all calculations presented in this paper are carried out in the flat-sky limit, it is straightforward to generalise to full sky, replacing the Fourier transforms with spherical harmonic transforms.

Finally, this approach could easily be extended to simulations with massive neutrinos, modified gravity or a hydrodynamical sector, such as to investigate possible degeneracies between the impact of distinct physical phenomena (i.e. IA, baryons, neutrinos and gravity) on higher-order lensing statistics.

\section{Conclusions}
\label{sec:conclusion}

Intrinsic alignments of galaxies cause a secondary signal in cosmic shear measurements that, if unaccounted for, will create catastrophic biases ($>$10$\sigma$) in the cosmology inferred from the next generation of lensing surveys. The physical mechanism of this effect is still under debate, however many models proposed in the literature seem to agree relatively well with current data \citep{Johnston_IA, Blazek2019, Fortuna2020}. While these are primarily aimed for two-point statistics, we show in this paper that different IA models can also be infused at the catalogue level, which can subsequently serve to model the IA on statistics beyond two-points.

We construct 2D tidal field maps with the projected mass sheets extracted from the {\it cosmo}-SLICS $w$CDM simulations, and couple these maps with lensing galaxy catalogues in order to assign an intrinsic alignment to individual objects. Our current model includes three free-parameters, namely the smoothing scale $\sigma_{\rm G}$, the tidal coupling parameter $A_{\rm IA}$ and the density-weighting parameter $b_{\rm TA}$. We validate our infusion model by comparing two-point statistics  against predictions from the NLA model (including the $GG$, $GI$, $II$ terms) and recover good match for most tomographic bins. The $II$ term is less well modelled, likely due to its increased sensitivity to projection effects.

We measure the impact of the $\delta$-NLA IA model on three different aperture mass map statistics -- peaks, minima and lensing PDF -- and find that region of large (positive or negative) \snr are more affected, especially when combining galaxies with different tomographic bins. Of these three probes, the lensing minima are more immune, primarily due to their lesser sensitivity to regions of high density (and therefore of high IA). 
When propagated onto a full cosmological inference, we find that the $\gamma$-2PCFs suffer the largest biases in presence of unmodelled IA. Table \ref{table:cosmo_bias_01} summarizes the cosmological biases incurred in presence of unmodelled IA, for various tomographic choices. Most notably, and for all probes, the inferred $w_0$ is shifted to more negative values, while $S_8$ and $\Omega_{\rm m}$ are lower compared to the input truth. 

Some sort of IA modelling will be required for all probes, however the effect is more localized than for $\gamma$-2PCFs, for which all angular scales are affected. In general the study presented here offers an avenue to identify parts of the non-Gaussian data vector that are the least affected by IA, given a specified IA model. We intend to proceed with a dedicated investigation of this in an upcoming work, notably for Minkowski functionals, voids and clipped shear. The method we develop here can be easily extended to marginalise over $A_{\rm IA}$ and $b_{\rm TA}$ for any weak lensing estimator, simply by multiplying the `$\epsilon^{\rm IA}_{1/2}$' catalogue entries by a real number and combining with the cosmic shear $\gamma_{1/2}$ columns. This opens up the possibility to jointly sample over the uncertain IA parameters in a joint [$\gamma$-2PCFs;$N_{\rm peaks}$(\snr\!\!)] weak lensing analysis, improving constraints on both the cosmological and the IA parameters.

\section*{Acknowledgements}

We would like to thank Simon Samuroff,  for sharing his density-to-tidal field conversion code, Jonathan Blazek for his insights on the TATT  model and on the interpretation of our measurements, and Ludovic van Waerbeke and Fran\c{c}ois Lanusse for useful discussions. We are thankful to the referee for pointing out an error in our equations (Eqs. 4-5-6), which after correction resulted in an improved agreement with the theoretical NLA model. JHD acknowledges support from an STFC Ernest Rutherford Fellowship (project reference ST/S004858/1). Computations were carried on the Cuillin cluster at the Royal Observatory of Edinburgh, which is managed by Eric Tittley, and on the Mardec cluster supported by the OCEVU Labex (ANR-11-LABX-0060) and the Excellence Initiative of Aix-Marseille University - A*MIDEX, part of the French ``Investissements d'Avenir'' program.  RR is supported by the European Research Council (Grant No. 770935).
\\
\\
\\
{\footnotesize All authors contributed to the development and writing of this paper. JHD led the analysis; NM conducted the cosmology inference, while RR co-developed the IA infusion formalism. }

\section*{Data Availability}

The SLICS numerical simulations can be found at http://slics.roe.ac.uk/, while the cosmo-SLICS can be made available upon request.



\bibliographystyle{hapj}
\bibliography{IA} 




\appendix

\section{Projected Tidal Field}
\label{sec:3D_2D}

We derive in this Appendix the equation for the projected tidal field that is used in the main text (Eq. \ref{eq:sij_2D}).
Starting from Eq. (\ref{eq:sij}), the  three-dimensional tidal field projected along the $z$-axis can be computed as:
 \begin{eqnarray}
 s_{\rm 2D}(\boldsymbol x_{\perp}) &=& \sum_{z} s_{ij} ({\boldsymbol x}) \nonumber\\
 &=& \sum_{z} \int {\rm d}^3{\boldsymbol k} \,\, {\rm exp}[{\rm i} {\boldsymbol{x k}}] \left[\frac{k_i k_j}{k^2} - \frac{1}{3}\right]\widetilde{\delta}(\boldsymbol k)\mathcal{G}(\sigma_{\rm G})\, \nonumber \\
 &=& \sum_{z} \!\! \int \!\! {\rm d}^3{\boldsymbol k}  \,\, {\rm exp}[{\rm i}xk_x + {\rm i}yk_y + {\rm i}zk_z] \left[\frac{k_i k_j}{k^2} - \frac{1}{3}\right]\widetilde{\delta}(\boldsymbol k)\mathcal{G}(\sigma_{\rm G})\,. \nonumber \\ 
 \end{eqnarray}
The sum over $z$ affects only the third term of the exponential function, hence using the identity $\int_{-a}^{a} {\rm exp}[{\rm i}zk_z] {\rm d}z = 2a({\rm sin}k_za)(k_za) = 2\pi \delta_{\rm D}(k_z)$ in the large $a$ limit, we can write:
 \begin{eqnarray}
 s_{\rm 2D}(\boldsymbol x_{\perp})  & = & 2\pi \int {\rm d}^2{\boldsymbol k_\perp}   \,\, {\rm exp}[{\rm i}xk_x + {\rm i}yk_y] \times  \nonumber \\
 & & \left[\frac{k_i k_j}{k^2} - \frac{1}{3}\right]\widetilde{\delta}(\boldsymbol k_{\perp}, k_z=0)\mathcal{G_{\rm 2D}}(\sigma_{\rm G}) \, .
 \end{eqnarray}
Note that the Dirac delta function, $\delta_{\rm D}(k_z)$, ensures that $k^2$ appearing in the denominator can be computed with $k_x$ and $k_y$ only, receiving no contribution from $k_z$. The quantity $\widetilde{\delta}(\boldsymbol k_\perp, k_z=0)$ can therefore be identified as the $x-y$ Fourier transform of the $z$-projected three-dimensional density field, $\delta_{2D}(\boldsymbol x_\perp)$. As a result, the projected tidal field, at position $\boldsymbol  x_\perp$, can be evaluated in Fourier space by solving:
  \begin{eqnarray}
\widetilde{s}_{\rm 2D}(\boldsymbol k_{\perp})  =  \sum_{z}  \widetilde{s}_{ij} ({\boldsymbol k}) =2\pi \left[\frac{k_i k_j}{k^2} - \frac{1}{3}\right]\widetilde{\delta}_{2D}(\boldsymbol k)\mathcal{G_{\rm 2D}}(\sigma_{\rm G}) \, ,
 \label{eq:sij_2D_app}
 \end{eqnarray}
for $ij = (xx, yy, xy)$, then inverse Fourier transformed to produce the three tidal maps, $s_{xx}$, $s_{yy}$ and $s_{xy}$. 
 
The prescription to assign an intrinsic alignment based on the projected tidal field, described in Eq. (\ref{eq:tidal_th}), involves the combinations $(s_{xx} - s_{yy})$ and $s_{xy}$, which therefore correspond to:
\begin{eqnarray}
 \widetilde{\epsilon}_1^{\rm IA} (\boldsymbol k_{\perp})  &\propto& \left(\frac{k_x^2 - k_y^2}{k^2} \right) \widetilde{\delta}_{\rm 2D}(\boldsymbol k_{\perp})\mathcal{G_{\rm 2D}}(\sigma_{\rm G})  \, ,\nonumber \\
 \widetilde{\epsilon}_2^{\rm IA} (\boldsymbol k_\perp)  &\propto& \left(\frac{k_x k_y}{k^2} \right) \widetilde {\delta}_{\rm 2D}(\boldsymbol k_\perp)\mathcal{G_{\rm 2D}}(\sigma_{\rm G}) 
\end{eqnarray}

Aside from the smoothing kernel, these are the same filters that are used for converting convergence maps to shear maps under the \citet[][KS hereafter]{KaiserSquires} inversion:
 \begin{eqnarray}
 \widetilde{\gamma_1} (\boldsymbol \ell)  = \left(\frac{k_x^2 - k_y^2}{k^2} \right) \widetilde {\kappa}(\boldsymbol \ell) \, , \hspace{1cm} \widetilde{\gamma_2} (\boldsymbol \ell)  = \left(\frac{k_x k_y}{k^2} \right) \widetilde {\kappa}(\boldsymbol \ell)
 \label{eq:gamma_12}
\end{eqnarray}
meaning that on can linearly combine the mass sheets with the correct coefficients and obtain intrinsic ellipticities from a normal KS inversion. This fact is exploited in \citet{Fluri2019} to generate pure IA convergence maps that are then combined with the cosmic shear convergence maps to create a contamination.

 \section{Stage-IV biases for different $A_{\rm IA}$ \& $\sigma_{\rm G}$ values}
\label{sec:2PCF_IA}

\begin{figure*}
\begin{center}
\includegraphics[width=3.5in]{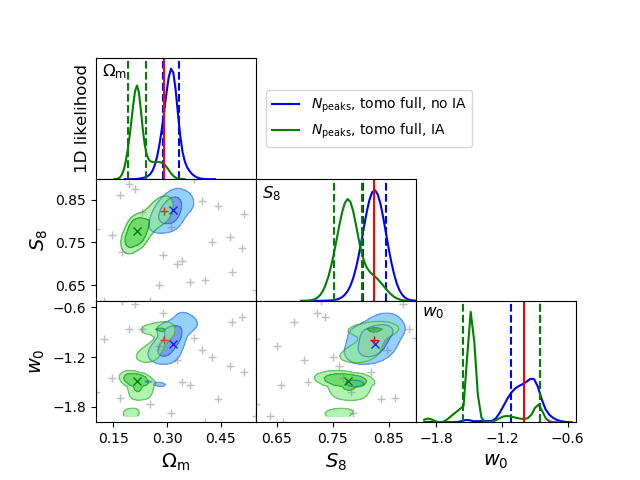}
\includegraphics[width=3.5in]{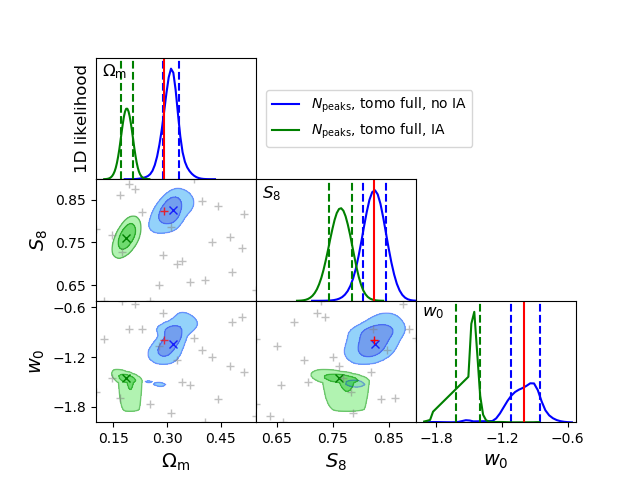}
\caption{Marginalized constraints from the full tomographic peak count statistics, with $A_{\rm IA}$= 1.0 ({\it left}) and 2.0 ({\it right}).}
\label{fig:like_varAIA}
\end{center}
\end{figure*}

\begin{figure*}
\begin{center}
\includegraphics[width=3.5in]{graphs/fig10d.png}
\includegraphics[width=3.5in]{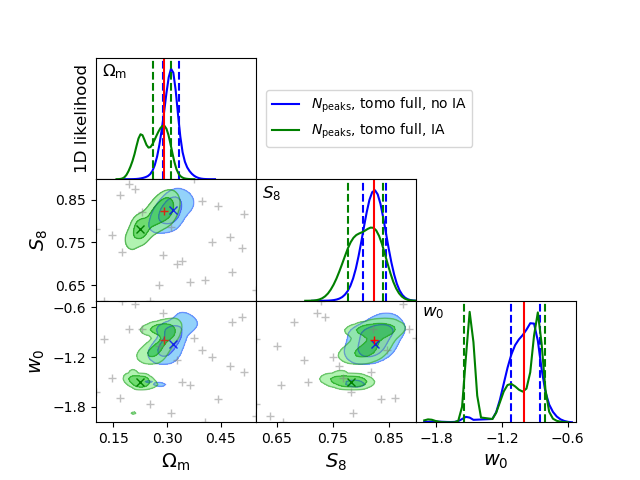}
\caption{Marginalized constraints from the full tomographic peak count statistics, with $\sigma_{\rm G}$=0.1$h^{-1}$Mpc ({\it left}) and $\sigma_{\rm G}$=0.5$h^{-1}$Mpc ({\it right}).}
\label{fig:like_varsmooth}
\end{center}
\end{figure*}

\begin{figure*}
\begin{center}
\includegraphics[width=3.5in]{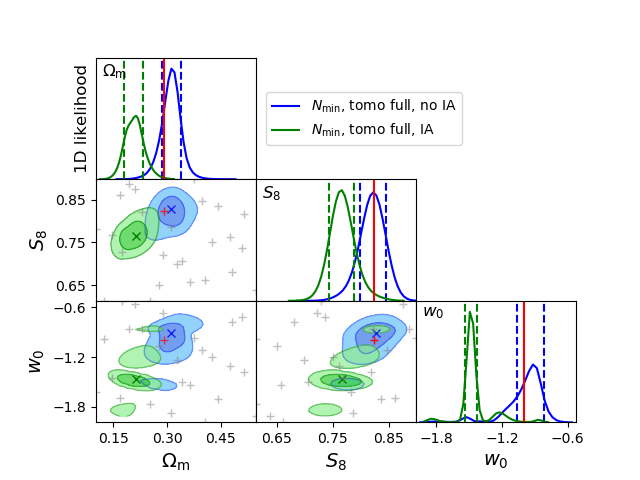}
\includegraphics[width=3.5in]{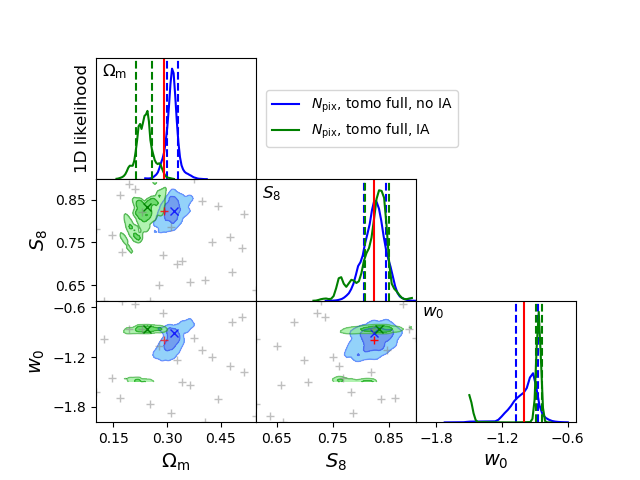}
\caption{Marginalized constraints from the full tomographic minima count statistics ({\it left}) and lensing PDF ({\it right}), with $A_{\rm IA}$= 1.5.}
\label{fig:like_varEST}
\end{center}
\end{figure*}

The main results in Sec. \ref{sec:Peaks} are reported cosmological biases for a coupling parameter set to $A_{\rm IA}$=1.5. In this Appendix, we extend these results to other analysis choices. We first show the marginalised constraints from the full tomographic peak statistics for $A_{\rm IA}$ = 1.0 and 2.0 in Fig. \ref{fig:like_varAIA}, the comparison between two smoothing scales $\sigma_{\rm G}$ (0.1 and 0.5 $h^{-1}$Mpc) in Fig. \ref{fig:like_varsmooth}, and the results from the three aperture mass map statistics in Fig. \ref{fig:like_varEST}. As expected, a stronger coupling with the tidal fields and a smaller smoothing length both enhance the cosmological biases. The marginalised posterior are tabulated in Table \ref{table:cosmo_bias_AIAvar} for different estimators and two coupling strengths, again assuming the $\delta$-NLA model. The choice of  smoothing scale has an important effect on all three parameters, and needs to be physically motivated, and possibly included as a free parameter in the model.

\begin{table*} 
\caption[]{Bias on the inferred cosmology caused by the intrinsic alignments of galaxies, if left unaccounted for, for the different cases listed in Table \ref{table:configuration}. Parameter shifts are reported for two different values of $A_{\rm IA}$. A smoothing scale of 0.1$h^{-1}$Mpc is assumed here. Numbers in parenthesis show the percentage of the marginalised error with respect to the input parameter value.}
\centering 
\begin{tabular}{clcccccc} 
\hline 
\hline
&& \multicolumn{3}{c}{$A_{\rm IA}=1.0$}& \multicolumn{3}{c}{$A_{\rm IA}=2.0$}\\
case& estimator(s)& $\Delta S_8$ & $\Delta w_0$ & $\Delta \Omega_{\rm m}$ & $\Delta S_8$ & $\Delta w_0$ & $\Delta \Omega_{\rm m}$ \\ 
\hline 
 (i)&$N_{\rm peaks}$, no tomo & -0.037 (-4.6\%) & 0.038 (3.8\%) & -0.007 (-2.2\%)  & -0.066 (-8.1\%) & 0.038 (3.8\%) & -0.007 (-2.2\%)\\
 (ii)&$N_{\rm peaks}$, tomo auto &  -0.003 (-0.3\%) & -0.009 (-0.9\%) & -0.01 (-3.4\%) & -0.006 (-0.7\%) & -0.009 (-0.9\%) & -0.015 (-5.2\%) \\
 (iii)&$N_{\rm peaks}$, tomo auto+pairs & -0.011 (-1.3\%) & 0.009 (0.9\%) & -0.028 (-9.6\%) & -0.049 (-6.0\%) & -0.6 (-60.0\%) & -0.113 (-39.0\%) \\
 (iv)&$N_{\rm peaks}$, tomo full & -0.049 (-5.9\%) & -0.545 (-54.5\%) & -0.095 (-32.6\%) &  -0.06 (-7.3\%) & -0.512 (-51.2\%) & -0.125 (-43.0\%) \\
\hline 
 (i)&$N_{\rm min}$, no tomo  & -0.037 (-4.4\%) & 0.0 (0.0\%) & 0.0 (0.0\%) & -0.073 (-8.9\%) & 0.0 (0.0\%) & 0.0 (0.0\%) \\
 (ii)&$N_{\rm min}$, tomo auto & 0.0 (0.0\%) & 0.0 (0.0\%) & 0.0 (0.0\%) &0.0 (0.0\%) & 0.0 (0.0\%) & 0.012 (4.0\%) \\
 (iii)&$N_{\rm min}$, tomo auto+pairs &-0.02 (-2.4\%) & -0.285 (-28.5\%) & -0.064 (-22.1\%) &-0.051 (-6.2\%) & -0.57 (-57.0\%) & -0.103 (-35.4\%)\\
 (iv)&$N_{\rm min}$, tomo full & -0.051 (-6.2\%) & -0.582 (-58.2\%) & -0.084 (-29.0\%) &  -0.064 (-7.7\%) & -0.564 (-56.4\%) & -0.116 (-39.8\%)  \\
\hline
 (i)&$N_{\rm pix}$, no tomo  & -0.046 (-5.6\%) & 0.037 (3.7\%) & 0.0 (0.0\%) & -0.078 (-9.5\%) & 0.027 (2.7\%) & 0.0 (0.0\%) \\
 (ii)&$N_{\rm pix}$, tomo auto &  -0.0 (-0.0\%) & 0.036 (3.6\%) & -0.015 (-5.3\%) & 0.003 (-0.3\%) & 0.036 (3.6\%) & -0.004 (-1.4\%) \\
 (iii)&$N_{\rm pix}$, tomo autopairs & 0.018 (2.2\%) & 0.039 (3.9\%) & -0.023 (-7.9\%) & 0.014 (1.7\%) & 0.039 (3.9\%) & -0.057 (-19.8\%) \\
 (iv)&$N_{\rm pix}$, tomo full &-0.006 (-0.8\%) & 0.059 (5.9\%) & -0.066 (-22.7\%) &   -0.025 (-3.0\%) & 0.04 (4.0\%) & -0.109 (-37.7\%)\\

\hline
\hline

\label{table:cosmo_bias_AIAvar}
\end{tabular} 
\end{table*}

\bsp	
\label{lastpage}
\end{document}

%% file: abbrev.tex
\let\jnlstyle=\rm
\def\refjnl#1{{\jnlstyle#1}}

\def\apjl{\refjnl{ApJ}}                
\def\apjs{\refjnl{ApJS}}               
\def\ao{\refjnl{Appl.~Opt.}}           
\def\aap{\refjnl{A\&A}}                

